\documentclass[prd,showpacs,showkeys,eqsecnum,nofootinbib,preprint,superscriptaddress]{revtex4-1}
\usepackage{epsfig}
\usepackage{amsmath}
\usepackage{amsfonts}
\usepackage{amssymb}
\usepackage{color}
\usepackage{graphicx}
\usepackage{url,hyperref}

\def  \bcen   {\begin{center}}
\def  \ecen   {\end{center}}
\def  \beq    {\begin{equation}}
\def  \eeq    {\end{equation}}
\def  \beqa   {\begin{eqnarray}}
\def  \eeqa   {\end{eqnarray}}

\def\bea{\begin{eqnarray}}
\def\eea{\end{eqnarray}}

\begin{document}
\title{ {Degenerate Higgs bosons decays to  ${\gamma\gamma}$ and ${Z\gamma}$ \\ in  the type II Seesaw Model}}
\author{M.~Chabab}
\email{mchabab@uca.ma}
\affiliation{Laboratoire de Physique des Hautes Energies et Astrophysique \\
D\'epartement de Physiques, FSSM, Universit\'e Cadi Ayyad, Marrakech, Morocco}
\author{M.~C. Peyran\`ere}
\email{michel.capdequi-peyranere@univ-montp2.fr}
\affiliation{Universit\'e Montpellier 2, Laboratoire Univers \& Particules de Montpellier - UMR 5299, F-34095 Montpellier, France}
\author{L.~Rahili}
\email{rahililarbi@gmail.com}
\affiliation{Laboratoire de Physique des Hautes Energies et Astrophysique \\
D\'epartement de Physiques, FSSM, Universit\'e Cadi Ayyad, Marrakech, Morocco}

\begin{abstract}
\noindent 
Using the most recent results of CMS and ATLAS, we study the Higgs decays to $\gamma\gamma$ and $Z\gamma$  in the scenario where the two CP even Higgs predicted by the type II seesaw model (HTM)  are close to mass degenerate with a mass near $125$ GeV.  We analyse the effects of the Higgs potential parameters constrained by the full set of perturbative unitarity,  boundedness from below (BFB) as well as  from precision electroweak measurements on these decay modes. Our analysis demonstrates that the observed excess in the diphoton Higgs decay channel can be interpreted in our scenario within a delineated region controlled by $\lambda_{1}$ and $\lambda_{4}$ coupling.  We also find a deviation in the Higgs decay to $Z\gamma$ with respect to the Standard Model prediction and the largest enhancement is found for a ratio $R_{Z\gamma}$ of the order $1.6$.  Furthermore we show that consistency with current ATLAS data on the diphoton decay channel favours a light doubly charged Higgs with mass in the range $92 - 180$~GeV.
Finally, we find that the $\gamma\gamma$ and $Z \gamma$  Higgs decay modes are generally correlated and the magnitude of correlation is sensitive to the sign of the $\lambda_{1}$ parameter.

\end{abstract}
\maketitle
%%%%%%%%%%%%%%%%%%%%%%%%%%%%%%%%%%%%%%%%
\section{Introduction}
The recent discovery of neutral scalar boson by  ATLAS \cite{atlas12} and CMS \cite{cms12} detectors at the Large Hadron Collider (LHC) corresponds  undoubtedly to the Higgs boson.
 All data collected at $7$ and $8$ TeV  supports the existence of Higgs signal with a mass around $125$ GeV with  Standard Model 
(SM) like properties. While Higgs production and decays into $W W$* and $Z Z$* are consistent with SM predictions, the data analyses still
 show a persisting deviation in the $\gamma\gamma$ channel for the gluon and vector boson fusion productions. It is interesting to consider
 theoretical contexts that can be reconciled with the current data, which means scenarios that realise enhancement in the diphoton Higgs decay.
 Such enhancement has recently been obtained in many studies based on SM model extensions which enlarge the Higgs sector with the 
inclusion of additional scalar charged states, providing a more involved and rich Higgs phenomenology.  Among these the extensions involving Higgs scalar triplets are particularly interesting well motivated models. The main motivation of the 
Higgs Triplet Models (HTM) is related to the neutrino mass generation which relies on seesaw mechanism \cite{mohapatra81}. Besides, a distinctive feature of HTM is the presence of exotic doubly charged state which can provide a clean and spectacular signature at colliders \cite{perez08, chiang10, akeroyd_Hpp}. On the other hand, it has been shown that the doubly charged Higgs played an important role to reconcile the $h \to \gamma\gamma$ enhancement observed at LHC \cite{aa12, chiang13}.\\

Recently, it has been emphasised that $\gamma\gamma$ like signal can be enhanced relative to the SM as a result of cohabitation of two nearly
 degenerate Higgs bosons at the observed 125 GeV. This possibility is particularly appealing and may be relevant as the experimental resolution
 cannot resolve yet the structure of two overlapping peaks. In \cite{gunion12}, the NMSSM has been used as prototype  model to show that the
 observed deviations from unity have strong potential to reveal the existence of almost degenerate resonances in the $\gamma\gamma$ signal. 
Also, in the context of 2HDM, two mass degenerate scalars consisting of a CP even Higgs $h$ and a CP odd Higgs $A$, can well reproduce 
 $\gamma\gamma$ enhancement without affecting the WW and ZZ signal \cite{ferreira12}. In this work, we extend previous analyses of the Higgs
 decay to diphoton in the type II seesaw \cite{aa12, akeroyd12, yagyu12, aoki13, chiang13, cschen13a, okada13, englert13, cschen13b, chchen14} to the scenario where the two CP even Higgs are nearly degenerate. A tentative to consider this situation in the Higgs Triplet Model (HTM) with $Y= 2$ has been done in \cite{akeroyd10, mariana12}. The former is one of the first to discuss the mass splitting between the two CP-even Higgs bosons and its dependence on their mixing angle.  More importantly, it studied the prospects for detection of these neutral higgses several search channels at the LHC in the maximal mixing scenario. The latter used a simplified formulation which only reproduces the results of unmixed neutral case providing suppressed relative branching ratio into $\gamma\gamma$ mode disfavoured by the current LHC measurements.   
 
This paper is organised as follows. In section 2, we review the main features of Higgs Triplet Model and present the full set of constraints 
on the parameters of the Higgs potential. Section 3 is devoted to the HTM scenario in the case where the two CP even Higgs are close to degenerate. 
 Here we perform a double analysis for Higgs decays to $\gamma\gamma$ and to $Z \gamma$. We summarise our main results in section 4. 

%**********************************************************
\section{Review of the type II Seesaw Model}\label{sec1}
\subsection{The Higgs Triplet Model}
Extension of the Higgs content by a real scalar triplet with
hypercharge $Y_\Delta=2$ implements the type II seesaw mechanism in the Standard Model. In this case  
the most general $SU(2)_{L}\times U(1)_{Y}$ gauge invariant scalar
potential for the Higgs Triplet Model ({\rm HTM}) reads as \cite{perez08, aa11} :
\begin{eqnarray}
V&=&-m_H^2{H^\dagger{H}}+\frac{\lambda}{4}(H^\dagger{H})^2+M_\Delta^2Tr(\Delta^{\dagger}{\Delta})\nonumber\\
&&+\lambda_1(H^\dagger{H})Tr(\Delta^{\dagger}{\Delta})+\lambda_2(Tr\Delta^{\dagger}{\Delta})^2
+\lambda_3Tr(\Delta^{\dagger}{\Delta})^2\nonumber\\
&&+\lambda_4{H^\dagger\Delta\Delta^{\dagger}H}+[\mu(H^T{i}\tau_2\Delta^{\dagger}H)+hc]
\label{higgspot} 
\end{eqnarray}
with $\Delta$ and $H$ are the Higgs triplet and doublet respectively, given by:
\begin{eqnarray}
\Delta &=\left(
\begin{array}{cc}
\delta^+/\sqrt{2} & \delta^{++} \\
\delta^0 & -\delta^+/\sqrt{2}\\
\end{array}
\right) \qquad {\rm and} \qquad H=\left(
                    \begin{array}{c}
                      \phi^+ \\
                      \phi^0 \\
                    \end{array}
                  \right)
                  \label{HDrep}
\end{eqnarray}

This general potential has 10 independent parameters
: two vev's ($v_d$, $v_t$), the $\mu$ parameter, five $\lambda's$, plus
$m_H^2$, and $M_\Delta^2$. We assume that all the potential 
parameters are real.

When the electroweak symmetry is spontaneously broken the Higgs doublet and Triplet 
fields acquire theirs vacuum expectation values.
\begin{equation}
\langle H \rangle = \frac{1}{\sqrt{2}} \left(
                    \begin{array}{c}
                      0 \\
                      v_d \\
                    \end{array}
                  \right), \qquad \langle \Delta \rangle = \frac{1}{\sqrt{2}}
\left(
\begin{array}{cc}

0 & 0 \\
v_t & 0\\
\end{array}
\right)
\label{HDvacuum}
\end{equation}
including the $W$ and $Z$ masses,
$m_W^2=g^2v^2/4$, with $v^2=(v_d^2+2v_t^2)\approx(246~\mathrm{GeV})^2$.
The minimization conditions that define the vacuum expectation values
in terms of the parameters of the potential are
\begin{eqnarray}
4m_H^2-\lambda v_d^2+4\sqrt{2}{\mu}v_t-2(\lambda_1+\lambda_4)v_t^2&=&0\label{tadpoles1}\\
2M_\Delta^2v_t-\sqrt{2}\mu{v_d^2}+(\lambda_1+\lambda_4)v_d^2v_t+2(\lambda_2+\lambda_3)v_t^3&=&0
\label{tadpoles2}
\end{eqnarray}
Explicitly, three of the ten degrees of freedom in the 
Higgs Triplet Model correspond to the three Goldstone bosons 
($G^\pm$, $G^0$) and the remaining seven become physical Higgs 
bosons, consisting of: three neutral scalars, $h^0$, $H^0$ (CP-even) and $A^0$ (CP-odd), which are mixtures of the neutral component of doublet and triplet fields, and a pair of charged  $H^\pm$ and 
$H^{\pm\pm}$ with masses $m_{h^0}$, $m_{H^0}$, $m_{A^0}$, $m_{H^\pm}$ 
and $m_{H^{\pm\pm}}$ respectively. 
%%%%%%%%%%%%%%%%%%%%%%%%%%%%%%%%
These symmetric matrices are diagonalised by the following two orthogonal matrices :
\begin{eqnarray}
{\mathcal{R}}_\alpha = \left(
\begin{array}{cc}
\cos\alpha & -\sin\alpha \\
\sin\alpha & \cos\alpha \\
\end{array}
\right),\,{\mathcal{R}}_\beta = \left(
\begin{array}{cc}
\cos\beta & -\sin\beta \\
\sin \beta & \cos\beta \\
\end{array}
\right)~{\rm and}~{\mathcal{R}}_{\beta^{'}} = \left(
\begin{array}{cc}
\cos\beta^{'} & -\sin\beta^{'} \\
\sin\beta^{'} & \cos\beta^{'} \\
\end{array}
\right)
\label{eq:rotamatalphabeta}
\end{eqnarray}
where the mixing angles $\beta,\,\beta'$ and $\alpha$  are given by :
\begin{equation}
\tan\beta=2\frac{v_t}{v_d},\,\,\tan\beta'=\sqrt{2}\frac{v_t}{v_d}\,\,{\rm and}\,\,\tan2\alpha=\frac{2 B}{A - C}.\label{eq:mixingangles}
\end{equation}
The coefficients $A, B$ and $C$ are the entries of the $CP_{even}$ mass matrix
\begin{equation}
{\mathcal{M}}_{CP_{even}}^2=
\left(
\begin{array}{cc}
A & B \\
B & C\\
\end{array}
\right)
\label{cpeven:matrix}
\end{equation}
where
\begin{equation}
A=\frac{\lambda}{2}v_d^2,\hspace{0.2cm}
B=v_d(-\sqrt{2}\mu+(\lambda_1+\lambda_4)v_t),\hspace{0.2cm}
C=\frac{\sqrt{2}\mu\,v_d^2+4(\lambda_2+\lambda_3)v_t^3}{2v_t}
\label{ABC:cpeven}
\end{equation}

We are thus left with
seven independent parameters; namely $\lambda$, $(\lambda_i)_{i=1,\ldots,4}$,
$\mu$, and $v_t$.  Equivalently, we can instead choose %%
$m_{h^0}, m_{H^0}, m_{A^0},
m_{H^\pm}, m_{H^{\pm\pm}}, v_t$, 
and  $\alpha $,  \label{parameters}  %%
as the seven independent parameters. One can easily relate the physical scalar 
masses and mixing angles from Eq.~(\ref{higgspot}) to the potential 
parameters, $\lambda$, $\lambda_i$, $\mu$ and $v_i$, and invert 
them to obtain $\lambda$, $\lambda_i$ and $\mu$ in terms of the physical 
scalar masses and the mixing angle $\alpha$ \cite{aa11}.

The seven independent parameters are usually chosen as 
$\lambda$, $\lambda_{i=1...4}$, $\mu$ and $M_\Delta$ (or $\lambda$, 
$\lambda_{i=1...4}$, $\mu$  and $v_t$). After using the minimisation conditions, 
the $10\times10$ squared mass matrix,
\begin{equation}
{\mathcal{M}}^2=\frac{1}{2}\frac{\partial^2V}{\partial^2{\eta_i^2}}\mid_{\Delta=\langle \Delta \rangle,H=\langle H \rangle}
\label{squaredmass}
\end{equation}
can be recast, using Eqs.~(\ref{tadpoles1}-\ref{tadpoles2}), in a block diagonal form of one 
doubly-degenerate eigenvalue $m_{H^{\pm\pm}}$ and four $2\times2$ matrices 
denoted by ${\mathcal{M}}_{\pm}^2$, ${\mathcal{M}}_{CP_{even}}^2$ 
and ${\mathcal{M}}_{CP_{odd}}^2$.
The masses of the various Higgses are given by :
\begin{eqnarray}
m_{H^{\pm\pm}}^2 & = & \frac{\sqrt{2}\mu{\upsilon_d^2}-\lambda_4\upsilon_d^2\upsilon_t-2\lambda_3\upsilon_t^3}{2v_t}\label{eq:mHpmpm}\\
m_{H^{\pm}}^2 & = & \frac{\upsilon^2[2\sqrt{2}\mu-\lambda_4\upsilon_t]}{4\upsilon_t}\label{eq:mHpm}\\
m_{A^0}^2 & = & \frac{\mu(\upsilon_d^2+4\upsilon_t^2)}{\sqrt{2}\upsilon_t}\label{eq:mA0}\\
m_{h^0}^2 &  = & \frac{A+C - \sqrt{(A-C)^2 + 4 B^2}}{2} \label{eq:mh0}\\
m_{H^0}^2 &  = & \frac{A+C +  \sqrt{(A-C)^2 + 4 B^2}}{2} \label{eq:mHH}
\end{eqnarray}
Notice that the CP even Higgs masses strongly depend on the values of $\lambda$ coupling  and $\mu$ parameter, while the mass of singly and doubly charged Higgs are only sensitive to the $\lambda_4$. \\

At this level we must stress that rather than assuming $\mu$ around the GUT scale, together with $\mu\simeq\,M_\Delta$, 
 we have found small values of $\mu$ which are consistent, at TeV scale, with a tiny value of $v_t$, necessary for realistic neutrino masses.
 The latter scenario for type II seesaw models is appealing since it is accessible to collider experiments. \\

\subsection{Theoretical and experimental constraints}
The HTM Higgs potential parameters are not free but have to obey  several constraints originating from theoretical requirements and experimental data. Thus any phenomenological studies are only reliable in the allowed parameter space. \\

First, recall that the LEP direct search results in the lower bounds  $m_{A^0, H^0} >  80-90$
~GeV for models with more than one doublet in the case of the neutral scalars. As to charged scalars triplets,  $m_{H^\pm, H^{\pm\pm}} > 80-100$~ GeV if the charged 
Higgs decays dominantly to leptons for general models with triplet. \\
Second, in HTM the $\rho$ parameter at level is given by the formula, $\rho \simeq 1 - 2 \frac{v_t^2}{v_d^2}$ which indicates a deviation from unity. Consistency with the current limit on $\rho$ from precision measurements \cite{db12} requires that the extra contribution $\delta\rho = - 2 \frac {v_t^2}{v_d^2}$ coming from Higgs scalar triplet should be negative and smaller than the limit $|\delta\rho| \leq
10^{-3}$.  At the 2 $\sigma$ level, the quoted number of $\rho$ parameter  
$\rho_0 = 1.0004^{+0.0029}_{-0.0011}$ \cite{db12} is 
well consistent with a negative $\delta \rho$. Moreover, 
relaxing the Higgs direct limit leads to 
$\rho_0 = 1.0008^{+0.0017}_{-0.0010}$, 
again compatible with $\delta \rho <0$. Further, we see that  
these experimental values place an upper limit on $v_t$ below $\leq 5$~GeV. \\

As to the theoretical constraints on the parameter space, we should take into account the perturbativity constraints
on the $\lambda_i$ as well as the stability of the electroweak vacuum that ensure that the potential is bounded from below (BFB). 

%%%%%%%%%%%%%%%%%%%%%%%%Unitarity & BFB%%%%%%%%%%%%%%%%%%%%%%%%%%%
For later use in the degenerate scenario, let us first recall the full set of constraints as obtained in \cite{aa11}: 

{\sl \underline{BFB}:}
\begin{eqnarray}
&& \lambda \geq 0 \;\;{\rm \&}\;\; \lambda_2+\lambda_3 \geq 0  \;\;{\rm \&}  \;\;\lambda_2+\frac{\lambda_3}{2} \geq 0
\label{eq:bound1} \\
&& {\rm \&} \;\;\lambda_1+ \sqrt{\lambda(\lambda_2+\lambda_3)} \geq 0 \;\;{\rm \&}\;\;
\lambda_1+ \sqrt{\lambda(\lambda_2+\frac{\lambda_3}{2})} \geq 0  \label{eq:bound2} \\
&& {\rm \&} \;\; \lambda_1+\lambda_4+\sqrt{\lambda(\lambda_2+\lambda_3)} \geq 0 \;\; {\rm \&} \;\; 
\lambda_1+\lambda_4+\sqrt{\lambda(\lambda_2+ \frac{\lambda_3}{2})} \geq 0 \label{eq:bound3}
\end{eqnarray}

{\sl \underline{Unitarity}:}
\begin{eqnarray}
&&|\lambda_1 + \lambda_4| \leq \kappa \pi  \label{eq:unit1} \\
&&|\lambda_1| \leq \kappa \pi \label{eq:unit2} \\
&&|2 \lambda_1 + 3 \lambda_4| \leq 2 \kappa \pi \label{eq:unit3} \\
&&|\lambda| \leq  2 {\kappa} \pi \label{eq:unit4} \\
&&|\lambda_2| \leq  \frac{\kappa}{2} \pi \label{eq:unit5} \\
&&|\lambda_2 + \lambda_3| \leq  \frac{\kappa}{2} \pi \label{eq:unit6} \\
&&|\lambda + 4 \lambda_2 + 8 \lambda_3 \pm \sqrt{(\lambda - 4 \lambda_2 - 8 \lambda_3)^2
+ 16 \lambda_4^2} \;| \leq  4 \kappa \pi \label{eq:unit7} \\
&&| 3 \lambda + 16 \lambda_2 + 12 \lambda_3 \pm \sqrt{(3 \lambda - 16 \lambda_2 - 12 \lambda_3)^2
+ 24 (2 \lambda_1 +\lambda_4)^2} \;| \leq  4 \kappa \pi \label{eq:unit8} \\
&&|2 \lambda_1 - \lambda_4| \leq 2 \kappa \pi \label{eq:unit9} \\
&&|2 \lambda_2 - \lambda_3| \leq  \kappa \pi \label{eq:unit10} 
\end{eqnarray}
where the parameter $\kappa$ takes the values $\kappa=8$ or $16$. \\
\noindent
These two sets of BFB and unitarity constraints can be reduced to a more compact 
analytical set where the allowed ranges for the $\lambda$'s are precisely identified: 
\begin{eqnarray}
&& 0 \leq \lambda \leq \frac{2}{3} \kappa \pi  \label{eq:syscompact1} \\
&& \lambda_2+\lambda_3 \geq 0  \;\;{\rm \&}  \;\;\lambda_2+\frac{\lambda_3}{2} \geq 0 \label{eq:syscompact2}\\
&& \lambda_2 + 2 \lambda_3 \leq \frac{\kappa}{2} \pi \label{eq:syscompact3}\\
&& 4 \lambda_2 + 3 \lambda_3 \leq  \frac{\kappa}{2} \pi \label{eq:syscompact4}\\
&& 2 \lambda_2 - \lambda_3 \leq  \kappa \pi  \label{eq:syscompact5}
\end{eqnarray}
and,
\begin{eqnarray}
&&  |\lambda_4| \leq 
 \min{\sqrt{(\lambda \pm 2 \kappa \pi) (\lambda_2 + 2 \lambda_3 \pm \frac{\kappa}{2} \pi)} } \label{eq:syscompact6}\\
&& |2 \lambda_1 + \lambda_4| \leq \sqrt{2 (\lambda - \frac{2}{3} \kappa \pi) (4 \lambda_2 + 3 \lambda_3 
- \frac{\kappa}{2} \pi)} \label{eq:syscompact7}
\end{eqnarray}

We stress here that these constraints are very restrictive conditions on the allowed range of the parameter space. All values
presented in the plots of our subsequent analysis are consistent with all theoretical and experimental bounds described in this section.
%%%%%%%%%% Degenerate Scenario%%%%%%%%%%%%%%%%%%%%
\section{Degenerate  Higgs Bosons Scenario in HTM}
In this section we will focus our analysis on the CP-even neutral Higgs bosons $h^0$ and $H^0$ with nearly degenerate mass.
First, we will study analytically several salient features of Higgs potential in this scenario. Particularly, emphasis will be put on $\lambda$, $\mu$, $v_t$ parameters, and on the mixing angle $\alpha$ and how their relations evolve near deneneracy. These relations would be very useful for subsequent phenomenological analysis. Then, we will see how the excess observed by LHC experiments in the Higgs decays to diphoton  and to $Z \gamma$ can be interpreted in our scenario. 

Other theoretical constraints modify in the degenerate case and new interesting relations among the Higgs potential parameters are derived in the appendix, specially the bounds on singly and doubly charged Higgs bosons.

\subsection{Analytic study}
From  the CP-even Higgs mass matrix given in Eq.~(\ref{cpeven:matrix}), the two eigenvalues $\lambda_{\pm}$ representing the squared masses of  $h^0$ and $H^0$, are defined as:
\begin{equation}
\lambda_{\pm} = \frac{A+C  \pm \sqrt{(A-C)^2 + 4 B^2}}{2}.
\label{dgnr1}
\end{equation}
where  $h^0$  is supposed to be lighter than $H^0$.

In this scenario, the difference of masses $\Delta M$ between the two neutral Higgs $H^0$ and $h^0$ is set to
about $2$~GeV, corresponding to the detector inability to resolve two nearly Higgs signals. If we note $M_{ex}$ the
 experimental Higgs boson mass,  then we have:
 $$ \lambda_+ - \lambda_- = (M_{H^0} - M_{h^0}) (M_{H^0} + M_{h^0}) \approx (M_{H^0} - M_{h^0}) 2 M_{ex} = 2 M_{ex} \Delta M.$$
So $\displaystyle{ \sqrt{(A-C)^2 + 4 B^2}} \leq 2 M_{ex} \Delta M$, which reduces to,
\begin{equation}
  |(\lambda_{1}+ \lambda_{4})v_t - \mu \sqrt 2 | \leq  \frac{M_{ex} \Delta M }{v_d}, 
  \label{dgnr2}
\end{equation}
when $A = C$. Numerically, the ratio  $\displaystyle{ \frac{M_{ex} \Delta M }{v_d} }$ is roughly equal to $1$~GeV for a Higgs mass about $125$~GeV.
 This tell us that the parameters  $\mu$ and $v_t$ should be of the same order. Besides, by setting  $A=C$ and for small $x= v_t/v_d$, the square root in the numerator of Eq.~(\ref{dgnr1}) simplifies to  $2 |B| = 2 v_d |(\lambda_{1}+ \lambda_{4}) v_t - \mu \sqrt 2|$, providing the following eigenvalues formula, 
\begin{equation}
\lambda_{\pm} = \frac{\lambda v_d^2}{4} + \frac{\mu \sqrt 2 v_d^2}{4 v_t} \pm v_d |(\lambda_{1}+ \lambda_{4}) v_t - \mu \sqrt 2 |.
\label{dgnr3}
\end{equation}
Now, we can also describe the CP-even neutral Higgs
 masses degeneracy as $m_{H^0} = M_{ex} (1 + H) $ and $m_{h^0} = M_{ex}(1 + h)$, where $H$ and $h$ are numerical parameters such 
that $(H-h) M_{ex}= \Delta M$. Hence $|H|$ and 
$\displaystyle{|h| \leq  \frac{\Delta M}{2 M_{ex}} \approx 8 \times 10 ^{-3}}$.
  Then $\lambda v_t - \mu \sqrt 2 = \displaystyle{2 v_t \frac{M_{ex}^2 - m_{A^0}^2 }{v_d^2}} + O(x)$ 
though the right formula is $\lambda v_t - \mu \sqrt 2 = 4 (\lambda_{2} + \lambda_{3}) v_d x^3$. 
 Similarly, for the CP odd Higgs mass, we write  $m_{A^0} = M_{ex}(1 + a)$, where $a$ denotes a numerical parameter. Then one gets:
$$\lambda v_t - \mu \sqrt 2 - 4 (\lambda_{2} + \lambda_{3}) v_d (v_t/v_d)^3 = 2 (-2 a + h + H) M_{ex} (v_t/v_d)^2  + O(v_t/v_d)^3.$$ 
This implies that $2 a \approx h + H$, which means that all the three neutral Higgs masses are close to the experimental mass $M_{ex}$
 with the following hierarchy $m_{H^0} \geq m_{A^0} \geq m_{h^0}$, and such that $m_{H^0}^2 + m_{h^0}^2 \approx 2 m_{A^0}^2$.\\ 

 Owing to these results, and within the approximation  where $v_t$ is much smaller than $v_d$, we easily derive the very useful relation,
 $\displaystyle{\frac{\mu}{\lambda}   \approx  \frac{v_t }{\sqrt 2} }$, which constrains the two potential parameters $\mu$ and  $ \lambda$. 
Also, for the present values of $M_{ex}$ and $v_d$,  the ratio  $ \displaystyle{\frac {\mu}{\lambda}} $ is approximately  equal to $ 0.375 $,  
which supports our previous remark that $v_t$ and $\mu$ have the same magnitude. 
It is also noticeable that the mass of the CP-odd neutral field, given by Eq.~(\ref{eq:mA0}),  
$m_{A^0}^2 = \displaystyle{ \frac{\mu (v_d^2 + 4 v_t^2)}{\sqrt 2 v_t}  }$  
reduces to $\displaystyle{M_{ex}^2 (1+  \frac{4 v_t^2}{v_d^2}) }$ for $v_t/v_d$ small, in agreement with the earlier analysis. This also means  
that the parameter $a$ is about $ \displaystyle{ \frac{2 v_t^2}{v_d^2}}$, two or three orders of magnitude less than the bounds on $H$ and $h$, 
suggesting that $H$ and $h$ should have different signs with the following hierarchy $m_{H^0} \geq M_{ex} \geq m_{h^0}$. 
Further, the two small parameters $H$ and $h$ can be recast in a more precise way as $ \displaystyle {H = 2 \frac{v_t^2}{v_d^2}  +
 \frac{\Delta M}{2 M_{ex}} }$ and  
$\displaystyle { h = 2 \frac{v_t^2}{v_d^2} - \frac{\Delta M}{2 M_{ex}} }$. \\
 Now we focus on the distinctive properties of the mixing angle $\alpha$ between the neutral components of the doublet and triplet Higgs fields.
We know that $\alpha$  is close to $\pi/4$ in the degenerate scenario since 
$\tan 2 \alpha$ is almost infinite. Moreover, from Eqs.~(7.22,7.23) \cite{aa11}, we have
$\displaystyle{\alpha \approx \pm \frac{\pi}{4} \pm 2 \frac{v_t}{v_d}}$ and   
 $\displaystyle{\sin \alpha \approx \pm \frac{1}{\sqrt 2} \pm \sqrt 2\frac{v_t}{v_d}}$, because the parameter 
$\bar \mu \approx \displaystyle{ \frac{2 \sqrt 2 m_A^2 v_t}{v_d^2} }$ $ \approx \mu $ for small $x= v_t/v_d$ as seen in Eq.~(\ref{eq:mA0}).
 The precise signs depend on the $\nu$ parameter, as defined in Eq.~(7.1) of \cite{aa11} and on $B$ given by Eq.~(\ref{ABC:cpeven}).  By assuming that $-\pi/2 <\alpha <\ pi/2$, we must also consider that:\\
$\bullet$ The sign of $B$ must be opposite to the sign of $\sin \alpha$, 
that is  $- \alpha$, as shown in \cite{aa11}. \\
$\bullet$ The sign of $\nu$ is relevant to determine the mixing angle near the degeneracy, Eq.~(7.23)  \cite{aa11}.\\
Indeed by writing $\displaystyle{\alpha_{\pm}(\epsilon) = \pm \frac{\pi}{4} -\epsilon 2 x}$,
 with $x=v_t/v_d$  and $\epsilon = \pm 1$, we find that: \\
$\bullet$ $B = - (m_{H^0}^2 - m_{h^0}^2) \cos \alpha \sin \alpha $ has the same sign of  $- \sin \alpha$.  Besides $B$ is approximated by
$B= \pm \displaystyle{\frac{(m_{H^0}^2 - m_{h^0}^2) \cos 4 x}{2} } $ which shows, for small $x$, that $B$ is positive for $\alpha_{-}$ and negative for $\alpha_{+} $. \\
$\bullet$ $\nu= \displaystyle{ \frac{m_{H^0}^2 - m_{h^0}^2}{2 x}} (4 x \cos 2 \alpha - \sin 2 \alpha )$ and  the sign of $\nu$ is the sign of $4 x \cos 2 \alpha - \sin 2 \alpha$.
\\ 
At this level, note that all these considerations are not sufficient to say more about the mixing angle $\alpha$. We postpone to the appendix a tentative to catch this angle owing to the BFB conditions Eqs.~(\ref{eq:bound1}-\ref{eq:bound3}).

%%%%%%%%%%%%%%%%%%%%%%%%%%%%%%%%%%%%%%%%%%%%%%%%%%%%%%%%%%%%%%%%%
\subsection{Phenomenological analysis}
In this subsection we explore the analysis of the Higgs decays to diphoton and and to $Z \gamma$  in HTM where the neutral Higgs states are nearly mass degenerate. \\
In the SM, the partial decay widths of scalar $h$ \cite{gunion2000} is given by : 
\begin{eqnarray}
\label{eq:SM-h2gaga}
\Gamma(h \rightarrow\gamma\gamma) = \frac{G_\mu\alpha^2 M_h^3}
{128\sqrt{2}\pi^3} \bigg| \sum_f N_c Q_f^2 A_{1/2}^h (\tau_f) +
 {A}_{1}^h (\tau_w) \bigg|^2
\label{width_gaga_sm}
\end{eqnarray}
with the amplitudes for spin-$\frac{1}{2}$ and spin-1 particles given by:
\begin{eqnarray}
A_{\frac{1}{2}}^{h}(\tau) &=& +2 [\tau +(\tau -1)f(\tau)]\, \tau^{-2}  \nonumber \\
A_1^{h}(\tau)  &= & -[2\tau^2 +3\tau+3(2\tau -1)f(\tau)]\, \tau^{-2}
\label{eq:A121scalar}
\end{eqnarray}
and the function $f(\tau)$ defined as,
\begin{eqnarray}
f(\tau)=\left\{
\begin{array}{ll}  \displaystyle
\arcsin^2\sqrt{\tau} & \tau\leq 1 \\
\displaystyle -\frac{1}{4}\left[ \log\frac{1+\sqrt{1-\tau^{-1}}}
{1-\sqrt{1-\tau^{-1}}}-i\pi \right]^2 \hspace{0.5cm} & \tau>1
\end{array} \right.
\label{eq:ftau}
\end{eqnarray}
In the HTM, this decay width becomes:
\begin{eqnarray}
\label{eq:HTM-h2gaga}
\Gamma({\mathcal{H}} \rightarrow\gamma\gamma)
& = & \frac{G_\mu\alpha^2 M_{{\mathcal{H}}}^3}
{128\sqrt{2}\pi^3} \bigg| \sum_f N_c Q_f^2 g_{{\mathcal{H}} ff} A_{1/2}^{{\mathcal{H}}}
(\tau_f) + g_{{\mathcal{H}} VV} A_1^{{\mathcal{H}}} (\tau_W) \nonumber \\
&& \hspace*{1.6cm} + {\tilde g_{{\mathcal{H}} H^\pm\,H^\mp}}A_0^{{\mathcal{H}}}(\tau_{H^{\pm}})+
4{\tilde g_{{\mathcal{H}} H^{\pm\pm}H^{\mp\mp}}}A_0^{{\mathcal{H}}}(\tau_{H^{\pm\pm}}) \bigg|^2
\label{width_gaga_sm}
\end{eqnarray}
where $\mathcal{H}$ is a generic notation for ${h^0}$  and ${H^0}$. The amplitudes $A_{1/2}^{{\mathcal{H}}},A_{1}^{{\mathcal{H}}}$ are
defined below, whereas $A_0^{{\mathcal{H}}}$ for spin-0 particle is
defined as \cite{djouadi2005gj},
\begin{eqnarray}
A_{0}^{{\mathcal{H}}}(\tau) &=& -[\tau -f(\tau)]\, \tau^{-2}
\label{eq:A0scalar}
\end{eqnarray}
$\tau_{i}=m^2_{\mathcal{H}}/4m^2_{i}$ $(i=f,w,H^{\pm},H^{\pm\pm})$
are the phases space functions. The reduced couplings $g_{\mathcal{H} ff}$
and $g_{\mathcal{H} VV}$ of the Higgs bosons to fermions and $W$ bosons are
given in Tab.\ref{table_couplings}, while the trilinear dimensionless couplings
${\tilde g_{{\mathcal{H}} H^{\pm\pm}H^{\mp\mp}}}$ and
${\tilde g_{{\mathcal{H}} H^\pm\,H^\mp}}$ to charged Higgs bosons
are related to the couplings in the Lagrangian,
${\mathcal{L}}= g_{{\mathcal{H}} H^{\pm}H^{\mp}}\mathcal{H}H^+H^-+g_{{\mathcal{H}} H^{\pm\pm}H^{\mp\mp}}\mathcal{H} H^{++}H^{--}+\dots$, as:
\begin{eqnarray}
\tilde g_{\mathcal{H} H^{++}H^{--}} & = &  -\frac{s_w}{e}\frac{m_W}{m_{H^{++}}^2}g_{\mathcal{H} H^{++}H^{--}}\\
\tilde g_{\mathcal{H} H^{+}H^{-}} & = &  -\frac{s_w}{e}\frac{m_W}{m_{H^+}^2}g_{\mathcal{H} H^+H^-}
\label{trilinear_coup1}
\end{eqnarray}
where the HTM trilinear couplings $g_{h^0 H^+H^-};g_{h^0 H^{++}H^{--}}$ are given for the light CP-even Higgs boson by:
\begin{eqnarray}
g_{h^0H^{++}H^{--}}&=&-\{2\lambda_2v_ts_\alpha+\lambda_1v_dc_\alpha\} \nonumber\\
g_{h^0H^+H^-}&=&-\frac{1}{2}\bigg\{\{4v_t\lambda_{23}^+c_{\beta'}^2+2v_t\lambda_1s_{\beta'}^2-
\sqrt{2}\lambda_4v_dc_{\beta'}s_{\beta'}\}s_\alpha \nonumber\\
&&+\{\lambda\,v_ds_{\beta'}^2+{\tilde \lambda_{14}^+}v_dc_{\beta'}^2+
(4\mu-\sqrt{2}\lambda_4v_t)c_{\beta'}s_{\beta'}\}c_\alpha\bigg\}
\label{trilinear_couph0}
\end{eqnarray}
with $\tilde \lambda_{14}^+=2 \lambda_1+ \lambda_4$, and $\tilde \lambda_{23}^+= \lambda_2+ \lambda_3$,
where for the heavy Higgs boson, these couplings are obtained simply from the above couplings by the substitutions 
\begin{eqnarray}
g_{H^0H^{++}H^{--}}&=&  g_{h^0H^{++}H^{--}}  [c_\alpha \rightarrow -s_\alpha, s_\alpha \rightarrow c_\alpha]
\label{eq:gHHpp}\\
g_{H^0H^+H^-}&=& g_{h^0H^+H^-}  [c_\alpha \rightarrow -s_\alpha, s_\alpha \rightarrow c_\alpha]
\label{eq:gHHp} 
\end{eqnarray}
\begin{table}[!h]
\begin{center}
\renewcommand{\arraystretch}{1.5}
\begin{tabular}{|c|c|c|c|c|} \hline\hline
\ \ $\mathcal{H}$ \ \  &$\tilde{g}_{\mathcal{H} \bar{u}u}$&
                  $\tilde{g}_{\mathcal{H} \bar{d}d}$&
$\tilde{g}_{\mathcal{H} W^+W^-} $ \\ \hline\hline
$h^0$  & \ $\; c_\alpha/c_{\beta'} \; $ \
     & \ $ \; c_\alpha/c_{\beta'} \; $ \
     & \ $ \; +e(c_\alpha\,v_d+2s_\alpha\,v_t)/(2s_W\,m_W) \; $ \ \\
$H^0$  & \ $\; - s_\alpha/c_{\beta'} \; $ \
     & \ $ \; - s_\alpha/ c_{\beta'} \; $ \
     & \ $ \; -e(s_\alpha\,v_d-2c_\alpha\,v_t)/(2s_W\,m_W) \; $ \ \\ \hline\hline
\end{tabular}
\end{center}
\caption{ The CP-even neutral Higgs couplings to 
fermions and gauge bosons in the HTM {\sl relative} to the SM Higgs couplings,
$\alpha$ and $\beta'$ denote the mixing angles respectively in the CP-even and
charged Higgs sectors, $e$ is the electron charge, $m_W$ the $W$ gauge boson
mass and $s_W$ the weak mixing angle \cite{aa12}.}
\label{table_couplings}
\end{table}
%
%%%%%%%%%%%%%%%%%%%%%%%%%%%%%%%%%%%%%%%Higgs to Z gamma%%%%%%%%%%%%%%%%%%%%%%%%%%%%%%%%%%%%%%%%%%%%%%%%%%%%%%%%%

For the Higgs decay into $Z\gamma$, the SM partial width is given by :
\begin{eqnarray}
\Gamma (h \to Z\gamma) &=& \frac{G^2_\mu M_W^2\,\alpha\,M_{h}^{3}}
{64\,\pi^{4}} \left(1-\frac{M_Z^2}{M_{h}^2} \right)^3 \bigg| \sum_{f} 
\frac{Q_f\, \hat{v}_f N_c}{c_W}\, {\cal F}^{h}_{1/2} 
(\tau_f,\lambda_f) + {\cal F}^{h}_1 (\tau_W,\lambda_W) \bigg|^2  \nonumber
\label{width_gaz_sm}
\end{eqnarray}
with now $\tau_i= 4M_i^2/M_H^2$, 
$\lambda_i = 4M_i^2 /M_Z^2$ and the amplitude factors are of the form : 
\begin{eqnarray}
{\cal F}_{1/2}^H (\tau,\lambda) & = & \left[I_1(\tau,\lambda) - I_2(\tau,\lambda)
\right] \\
{\cal F}_{1}^H (\tau,\lambda) & = & c_W \left\{ 4\left(3-\frac{s_W^2}{c_W^2} \right)I_2(\tau,\lambda) + \left\{ \left(1+\frac{2}{\tau}\right) \frac{s_W^2}{c_W^2} - \left( 5+\frac{2}{\tau} \right)\right\} I_1(\tau,\lambda)\right\}
\label{eq:hzgaform}
\end{eqnarray}
with $\hat{v}_f=2I_f^3-4 Q_f s_W^2$. The functions $I_1$ et $I_2$ 
are given by :
\begin{eqnarray}
I_1(\tau,\lambda) &=& \frac{\tau\lambda}{2(\tau-\lambda)}
+ \frac{\tau^2\lambda^2}{2(\tau-\lambda)^2} \left[ f(\tau^{-1})-f(\lambda^{-1}) 
\right] + \frac{\tau^2\lambda}{(\tau-\lambda)^2} \left[ g(\tau ^{-1}) - 
g(\lambda^{-1}) \right]\\
I_2(\tau,\lambda) &=& - \frac{\tau\lambda}{2(\tau-\lambda)}\left[ f(\tau
^{-1})- f(\lambda^{-1}) \right]
\label{I2function}
\end{eqnarray}
These functions can be expressed in terms of three-point Passarino-Veltman scalar functions \cite{pave1979} as : 
\begin{eqnarray}
C_{0,2}(m^2)\equiv C_{0,2}(M_Z^2,0,M_{\mathcal{H}}^2,m,m,m)\propto\frac{1}{m^2}I_{2,1}(\tau,\lambda)\nonumber
\end{eqnarray}
\noindent
where $C_0$ and $C_2$ are scalar integrals given often by 
$C_2\equiv C_{11}+C_{23}$ (see Ref. \cite{Hooft1979}).
the function $f(\tau)$ is defined above in Eq.~(\ref{eq:ftau}) whereas the $g(\tau)$ function can be expressed as 
\begin{equation}
g(\tau) = \left\{ \begin{array}{ll}
\displaystyle \sqrt{\tau^{-1}-1} \arcsin \sqrt{\tau} & \tau \ge 1 \\
\displaystyle \frac{\sqrt{1-\tau^{-1}}}{2} \left[ \log \frac{1+\sqrt{1-\tau
^{-1}}}{1-\sqrt{1-\tau^{-1}}} - i\pi \right] & \tau  < 1
\end{array} \right.
\label{eq:gtau}
\end{equation}
In the HTM, the charged contributions to $\mathcal{H} \to Z\gamma$ decay width reads as :
\begin{eqnarray}
\Gamma ({\cal H} \to Z\gamma) &=& \frac{G^2_\mu M_W^2\,\alpha\,M_{\cal H}^{3}}
{64\,\pi^{4}} \left(1-\frac{M_Z^2}{M_{\cal H}^2} \right)^3 \bigg| \sum_{f} 
\frac{Q_f\, \hat{v}_f N_c}{c_W} \,g_{\mathcal{H}ff}\, {\cal F}^{\cal H}_{1/2} 
(\tau_f,\lambda_f) \nonumber\\
&+&g_{\mathcal{H}WW}\,{\cal F}^{\cal H}_1 (\tau_W,\lambda_W)\\ 
&+&g_{ZH^+H^-}\tilde{g}_{\mathcal{H} H^\pm\,H^\mp}\,{\cal F}^{\cal H}_0 (\tau_{H^{\pm}},\lambda_{H^{\pm}}) 
+ g_{ZH^{\pm\pm}H^{\mp\mp}}\tilde{g}_{\mathcal{H} H^{\pm\pm}\,H^{\mp\mp}}\,{\cal F}^{\cal H}_0 (\tau_{H^{\pm\pm}},
\lambda_{H^{\pm\pm}}) \bigg|^2\nonumber
\label{width_gaz_htm}
\end{eqnarray}
 the ${\cal F}^{\cal H}_0 (\tau_{H^{\pm}},\lambda_{H^{\pm}})$, ${\cal F}^{\cal H}_0 (\tau_{H^{\pm\pm}},\lambda_{H^{\pm\pm}})$ factors reflecting the charged contributions
for $\Gamma_{\gamma\,Z}$ can be read in terms of the function $I_1(\tau,\lambda)$ previously defined as follows:
\begin{equation}
\label{fhp_fhpp}
{\cal F}^{\cal H}_0 (\tau_{H^{\pm}},\lambda_{H^{\pm}}) = 2 I_1(\tau_{H^{\pm}},\lambda_{H^{\pm}}) \quad ,\quad {\cal F}^{\cal H}_0 (\tau_{H^{\pm\pm}},\lambda_{H^{\pm\pm}}) = 4 I_1(\tau_{H^{\mp\pm}},\lambda_{H^{\mp\pm}}) 
\end{equation}

The reduced couplings for the CP-even Higgs bosons and the charged Higgs are as above, where the $g_{ZH^{\pm}H^{\mp}}$, $g_{ZH^{\pm\pm}H^{\mp\mp}}$  trilinear couplings can be expressed as :
\begin{eqnarray}
g_{ZH^{\pm}H^{\mp}}  & = & -\frac{1}{2}[-(c_w^2\,s_{\beta^{'}}^2)+(2c_{\beta^{'}}^2+s_{\beta^{'}}^2)\,s_w^2]/(s_w\,c_w)\\
g_{ZH^{\pm\pm}H^{\mp\mp}} & = & +[1-2\,s_w^2]/(s_w\,c_w)
\label{eq:couplagesZHpZHpp}
\end{eqnarray}
It is obvious that the contributions of the $H^{\pm\pm}$ and $H^{\pm}$ loops depend
on the details of the scalar potential.
The phase space functions $A_0$ and ${\cal F_0}$ involve the scalar masses
$m_{\mathcal{H}}$, $m_{H^\pm}$, and $m_{H^{\pm\pm}}$,
while $g_{\mathcal{H} H^+H^-};g_{\mathcal{H} H^{++}H^{--}}$ are sensitive to
several Higgs potential parameters. Since the interference between the charged scalar loop contributions
 and  those of the $W^\pm$ and $f(=t,b,c,\tau)$
loops depend on the sign of $g_{\mathcal{H}H^+H^-}$ and
$g_{\mathcal{H} H^{++}H^{--}}$, one would achieve either an enhancement or suppression of the $\mathcal{H}\to\gamma\gamma, Z\gamma$ decay modes with respect to SM predictions. \\

Hereafter we focus our analyse on the scenario where the two CP-even Higgs bosons $h^0$ and $H^0$ are nearly mass degenerate. In this scenario,
 the charged Higgs boson loops are included and the $g_{\mathcal{H}ww},  g_{\mathcal{H}\bar{f}f}$ couplings given in Tab.\ref{table_couplings} are used. 
 In order to infer limits on the parameters of our model from the experimental searches, we define the signal strengths by the ratios of the observed $\sigma^{\gamma\gamma, Z \gamma}/\sigma^{\gamma\gamma, Z \gamma}_{SM}$ and compare it to the following quantities :\\ 
\begin{eqnarray}
R_{\gamma\gamma} &=& R_{\gamma\gamma}(h^0)+R_{\gamma\gamma}(H^0) =
 \frac{\Gamma_{h^0\to\,gg}^{HTM}\times\,BR_{h^0\to\gamma\gamma}^{HTM}+\Gamma_{H^0\to\,gg}^{HTM}\times\,
BR_{H^0\to\gamma\gamma}^{HTM}}{\Gamma_{\Phi\to\,gg}^{SM}\times\,BR_{\Phi\to\gamma\gamma}^{SM}} \label{eq:RgagaTOT}\\
R_{Z\gamma} &=& R_{Z\gamma}(h^0)+R_{Z\gamma}(H^0) = \frac{\Gamma_{h^0\to\,gg}^{HTM}\times\,
BR_{h^0\to\,Z\gamma}^{HTM}+\Gamma_{H^0\to\,gg}^{HTM}\times\,BR_{H^0\to\,Z\gamma}^{HTM}}{\Gamma_{\Phi\to\,gg}^{SM}\times\,
BR_{\Phi\to\, Z\gamma}^{SM}}\label{eq:RgaZTOT}
\end{eqnarray}
\\
which require to compute the production cross section times the Higgs branching ratios to $\gamma\gamma$ and $Z \gamma$ relative to their corresponding values for the SM Higgs boson. The computation are performed by interfacing Feynrules with FormCalc. The current signal strength in the $pp \to \mathcal{H} \to\, \gamma\gamma$ channel is $1.57 ^{+0.33}_{-0.28}$ (ATLAS) \cite{atlas_rgg} and $0.78 ^{+0.28}_{-0.26}$  (CMS) \cite{cms_rgg}. Although CMS measurements is consistent with the Standard Model expectation within $1 \sigma$, ATLAS still observes almost $2 \sigma$ excess in this channel. Furthermore, we can see that the reported errors are still quite large. Therefore, it is worth to investigate whether the excess in the diphoton channel along with the wide range of $R_{\gamma\gamma}$ can be interpreted in the scenario where both CP- even Higgs bosons $h^0$ and $H^0$ appear in the $\gamma\gamma$ signal, with $M_{h^0} \approx  M_{H^0}$.  \\

As to $pp \to \mathcal{H} \to\, Z\gamma $ channel, no excess has been observed by LHC experiments and the current analyses provide only upper limits on the signal strengths, $\mu \leq 11$ (ATLAS) \cite{atlas_rzg} while $\mu \leq 9$ (CMS)  \cite{cms_rzg}. \\
%%%%%%%%%%%%%%%%%%%%Numerical analysis%%%%%%%%%%%%%%%%%%%%%%%%%%%%%%%%%%%%%%%%%%%%%%%%%%%%%%%%%%%%%%%%%%%%%%%%%%%%%%%%%%%%%%

In our numerical evaluation we use the following experimental input parameter:
$G_F= 1.166 \times 10^{-5}$~GeV$^{-2}$, $\alpha^{-1} \approx 128$, $m_Z=91.1875$~GeV, $m_W=80.45$~GeV, $m_t=173.39$~GeV and $M_{H}=125$~GeV with $ \Delta M \leq 2$~GeV. The Higgs mass fixes the Higgs self coupling to  $\lambda=0.53$. The total width of the Higgs boson is computed  including leading order  QCD  corrections as given in \cite{djouadi1996} as well as the  off-shell decays $\mathcal{H}\to WW^*$ and $\mathcal{H}\to ZZ^*$ \cite{Rizzo1980gz}.     
We also assume $v_t = 1$~GeV and $\mu=0.37 - 0.4$~GeV, these values satisfy the condition $\displaystyle{\frac{\mu}{\lambda}   \approx  \frac{v_t }{\sqrt 2} }$ resulting from the nearly mass degeneracy of  two CP-even Higgs bosons as discussed in Sec. III.  \\

For the singly charged Higgs mass we use the LEP II latest bound, $m_{H^{\pm}} \geq 88$~GeV \cite{LEP} as well as the limits established by the LHC $m_{H^{\pm}}$  $\leq 666$~GeV \cite{atlas_charged, cms_charged}. 
In the case of the doubly charged Higgs masses, we take into account the recent experimental upper limits $m_{H^{\pm \pm}} \geq 409$~GeV \cite{atlas_dcharged} and $m_{H^{\pm \pm}} \geq 445$~GeV  \cite{cms_dcharged}, reported by ATLAS and CMS respectively, assuming 100\% branching ratio for $H^{\pm\pm}\to l^\pm l^\pm$ decay. Notice that in realistic cases one can easily find scenarios where this decay channel is suppressed with respect to  $H^{\pm\pm}\to W^{\pm} W^{\pm (*)}$  \cite{akeroyd77, garayoa08, kadastik08} which could invalidate partially the LHC limits. In HTM analysis with relatively large triplet' VEV,  $v_t \approx 1$~GeV, the $H^{\pm\pm}\to W^{\pm} W^{\pm *}$ decay channel can still overpass the two-sign same lepton channel for $m_{H^{\pm\pm}}$ and the limit goes down all the way to $100$~GeV \cite{goran12, chiang12, shinya13, re13}. We will consider this value as a nominal lower bound in our subsequent numerical discussion. \\
 
As discussed above the analysis hereafter focuses on the most relevant potential parameters ($\lambda_1$, $\lambda_4$) taking into account the unitarity and BFB constraints  as well as the relations on potential parameters resulting from mass degeneracy of the neutral Higgs bosons. First, from Eqs.(\ref{trilinear_couph0}-\ref{eq:gHHp}) notice that the Higgs couplings to the scalar triplet $H^{\pm}$ and $H^{\pm \pm}$ are sensitive to $\lambda_1$ parameter. Variation of $\lambda_1$, with fixed $\lambda_4$, enhances (reduces) these trilinear couplings ($g_{\mathcal{H}H^\pm H^\mp}$ and  $g_{\mathcal{H}H^{\pm\pm} H^{\mp\mp}}$), leading to a destructive (constructive) interference between loop contributions. One can also see  that  the magnitude of $H^{\pm \pm}$ contribution is more important than $H^\pm$ one, due to a factor $4$ coming from its doubled electric charge and to its reduced mass in most of the parameter space where $R_{\gamma\gamma, Z\gamma}$ is enhanced. Furthermore since $m_{H^\pm}$ and $m_{H^{!
 \pm\pm}}$ contributions also depend on  $\lambda_4$, we find that $m_{H^{\pm}} \geq m_{H^{\pm \pm}}$ when $\lambda_4$ is positive, while this hierarchy is reversed for negative $\lambda_4$. 
In the former case, the enhancement ( $R_{\gamma\gamma, Z\gamma} \geq 1$) reduces $H^{\pm\pm}$ and $H^\pm$ leading to boosted couplings that scale like the inverse second power of these masses. To illustrate this point, we show that varying $\lambda_4$ between $-3$ and $1$ decreases $m_{H^{\pm \pm}}$ from $328$ to $92$ ~GeV, while for $-1 <  \lambda_4 < 0$  the $m_{H^{\pm \pm}}$ reduces to smaller values in the range $216$ to $127$~GeV.\\

\begin{figure}[ht]
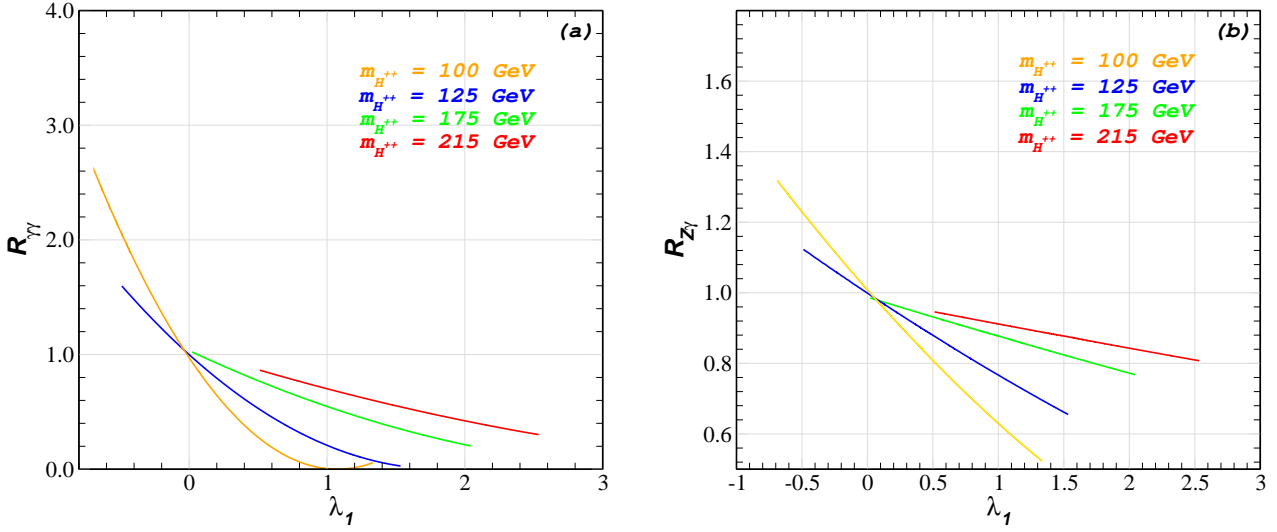

\hspace{-0.5cm}
\begin{tabular}{rr}
\hspace{-.8cm}\resizebox{80mm}{!}{\includegraphics{dgnr_Rgaga_ld1_new.eps}}&
\hspace{.6cm}\resizebox{80mm}{!}{\includegraphics{dgnr_RgaZ_ld1_new.eps}}
\end{tabular}
\caption{The $R_{\gamma\gamma}$ (left) and $R_{Z\gamma}$ (right) ratios as a function of $\lambda_1$ for various values of  $m_{H^{\pm\pm}}=\{100, \,125 ,\,175, \, 215\}$~GeV with $-5 \le \lambda_2 \le 5$ and $\lambda_3 = 2 \lambda_2$.}
\label{fig:Rxx_ld1}
\end{figure}

We plot in Fig.~\ref{fig:Rxx_ld1} $R_{\gamma\gamma}$ and $R_{Z\gamma}$, defined in Eqs.~(\ref{eq:RgagaTOT}-\ref{eq:RgaZTOT}),  as a function of  $\lambda_1$ for various values of $m_{H^{\pm\pm}}$. Since the $H^{\pm \pm}$ and $H^{\pm}$ contribution are shutdown for vanishing  $\lambda_1$, the trilinear couplings being very small, the SM contributions are dominant In this case.  When $\lambda_1$ moves to small values centred at zero, with a relatively light charged Higgses, a large enhancement is achieved in the $H \to \gamma\gamma$ decay channel leading to a magnitude of $R_{\gamma\gamma}$ that can goes up to $3.5$. The current ATLAS data on this decay mode favours a light doubly charged Higgs with mass $m_{H^{\pm \pm}} \leq 180$~GeV. Similar behaviour is seen for the $\mathcal{H} \to Z\gamma$ decay channel  (Fig.~\ref{fig:Rxx_ld1}-b) but in this case the deviation is very slight and the magnitude of $R_{Z\gamma}$ do not exceeds $60\%$ with respect to the SM prediction. \\

\begin{figure}[ht]
\hspace{-.95cm}
\begin{tabular}{rr}
\hspace{-.8cm}\resizebox{86mm}{!}{\includegraphics{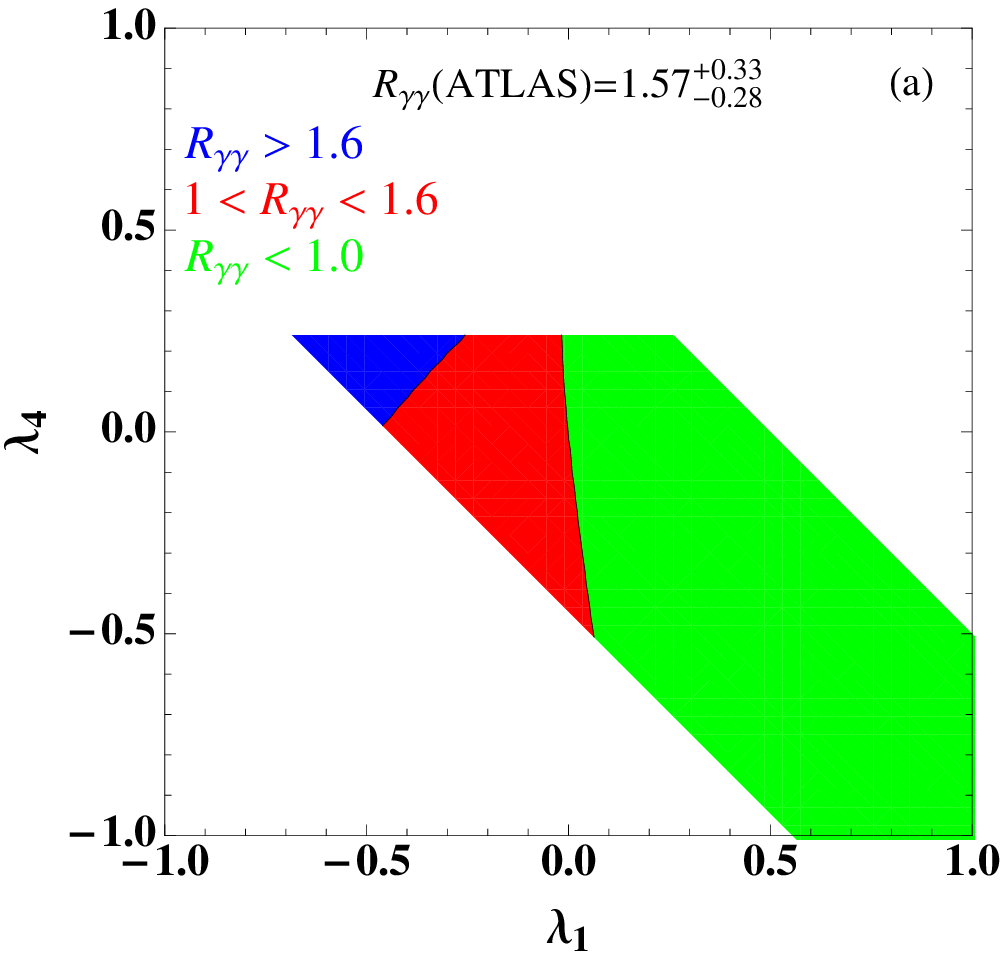}}&
\hspace{.1cm}\resizebox{86mm}{!}{\includegraphics{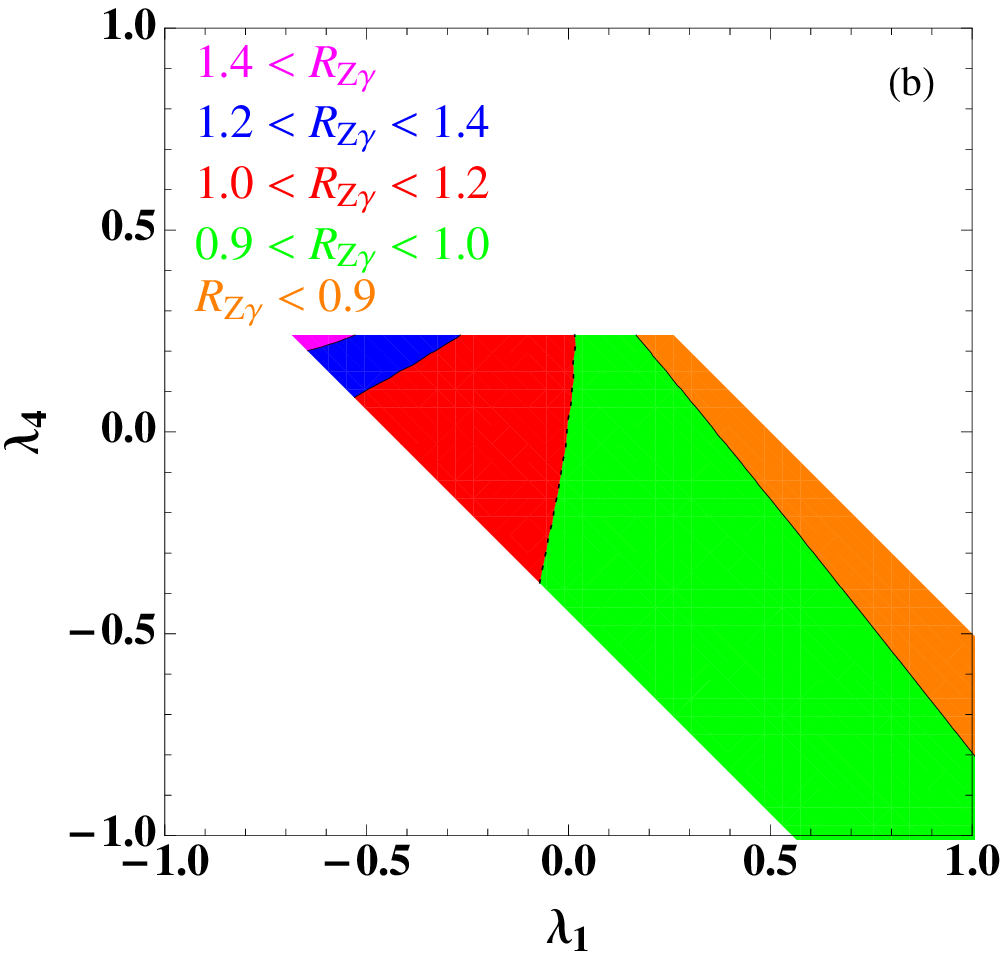}}
\end{tabular}
\caption{Scatter plots showing the $R_{\gamma\gamma}$ (left) and $R_{Z\gamma}$ (right) ratios in the $[\lambda_1,\lambda_4]$  with  $-1 \le \lambda_1 \le 3$, $-5 \le \lambda_2 \le 5$, $\lambda_3 = 2 \lambda_2$ and $-3 \le \lambda_4 \le 1$. The charged Higgs bosons masses varie as $110\,{\rm ~GeV}\le m_{H^\pm} \le 248\,{\rm ~GeV}$ and $92\,{\rm ~GeV}\le m_{H^{\pm\pm}} \le 328\,{\rm ~GeV}$.}
\label{fig:contour_ld1_ld4}
\end{figure}

Illustrating the behaviour seen in previous analysis we show in Fig.~\ref{fig:contour_ld1_ld4} a scatter plot in the $(\lambda_1,\lambda_4)$ plane for $R_{\gamma\gamma}$ and $R_{Z\gamma}$ ratios with variation of the singly (doubly) charged Higgs masses in the range [$110, 240$]~GeV corresponding to $-3 < \lambda_4 < 1$ ([$92, 328$]~GeV) respectively. Clearly one can delineate the regions of the parameter space where $R_{\gamma\gamma}$ deviates from the Standard Model. The figure in the right panel demonstrates that significant enhancement is achieved when both $\lambda_1$ and $\lambda_4$ move towards small values and shows explicitly the relevant regions of $(\lambda_1,\lambda_4)$ which are consistent with current ATLAS measurement within $1\sigma$ (red) and $2 \sigma$ (green).  For instance,  one can see that agreement with ATLAS diphoton data at $2\sigma$ requires their values to lie in the ranges  [$-0.6, 0.1$] and [$-0.5, 0.25$] respectively.\\ 

\begin{figure}[ht]
\hspace{-0.9cm}
\begin{tabular}{rr}
\hspace{-.8cm}\resizebox{86mm}{!}{\includegraphics{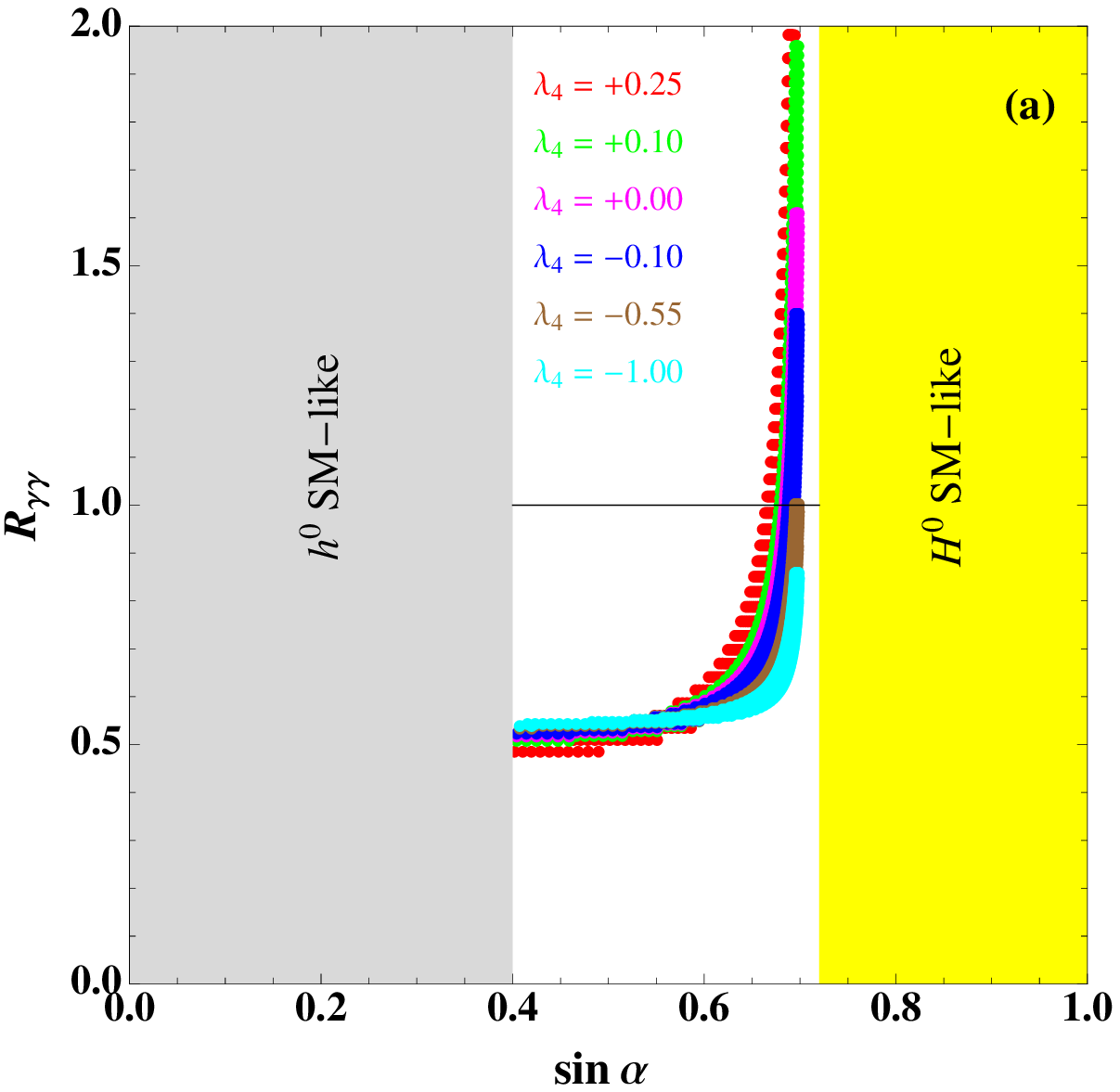}}&
\hspace{.1cm}\resizebox{86mm}{!}{\includegraphics{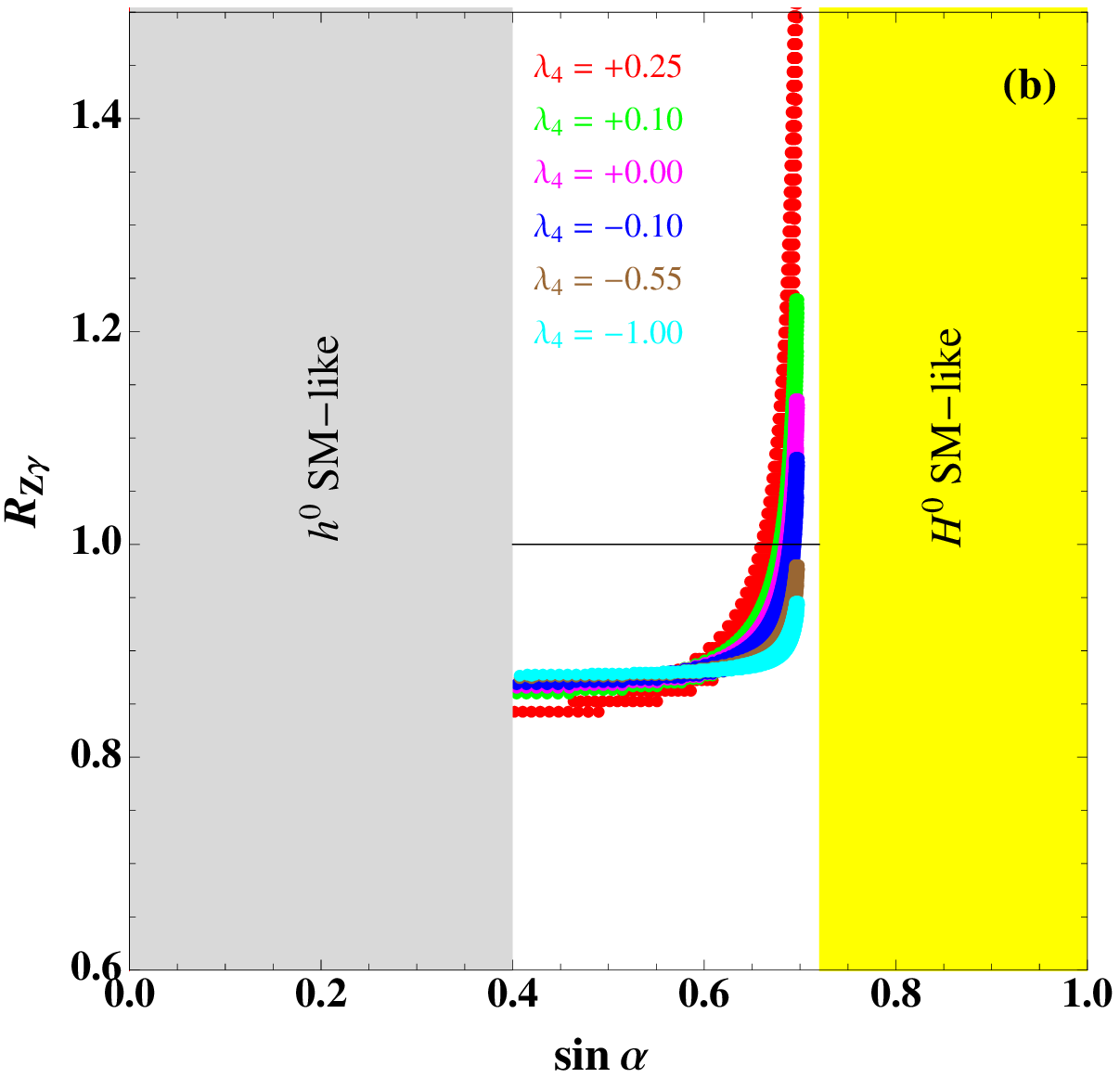}}
\end{tabular}
\caption{Scatter plots showing the $R_{\gamma\gamma}$ (left) and $R_{Z\gamma}$ (right) as a function of $\sin\alpha$ for various values of $\lambda_4$ with $-1 \le \lambda_1 \le 3$, $-5 \le \lambda_2 \le 5$ and $\lambda_3 = 2 \lambda_2.$.}
\label{fig:Rxx_sina}
\end{figure}

In Fig.~\ref{fig:Rxx_sina} we plot the ratios versus the absolute value of $\sin\alpha$ for various values of $\lambda_4$. As one can appreciate our results are relatively sensitive to variation of this mixing angle within the interval [$0.4, 0.75$] corresponding to the nearly mass degenerate scenario. The grey (yellow) region represents the non degenerate case where $h_0$ ($H_0$) is the SM like Higgs respectively. This figure shows a suppression in both decay modes with almost constant ratios, $R_{\gamma\gamma}<0.5$ and $R_{Z\gamma} < 0.9$, for most of $\sin\alpha$ range ($0.4<\sin\alpha<0.65$), except for very narrow interval in the vicinity of $\sin\alpha \approx 0.75$ where  $R_{\gamma\gamma, Z\gamma}$ drop quickly exceeding SM values. We also note that $R_{\gamma\gamma, Z\gamma}$ are almost unaffected by $\lambda_4$ when $\lambda_4 \geq 0.25$, but a significant (moderate) enhancement of $H \to \gamma\gamma$ ($H \to Z\gamma$) is reached for $\lambda_4$ getting small negative values in the interval [$-0.5, 0.25$] which are equivalent to light $H^{\pm \pm}$ bosons with mass bellow $215$~GeV.
These results disagree with the analysis performed by Franck et al. \cite{mariana12} for degenerate case where they found a suppressed $\gamma\gamma$ signal for  $\sin\alpha$ lying anywhere between $0$ to $1$. The latter is a trivial scenario since it reproduces exactly the unmixed case ($\sin\alpha = 0$), furthermore it is incompatible with LHC measurements. In the case of $\mathcal{H} \to Z\gamma$ decay channel, they found that $R_{Z\gamma}$ is almost insensitive to $\sin\alpha$ with some modest enhancement for small mixing $0< \sin\alpha< 0.2$. \\

\begin{figure}[ht]
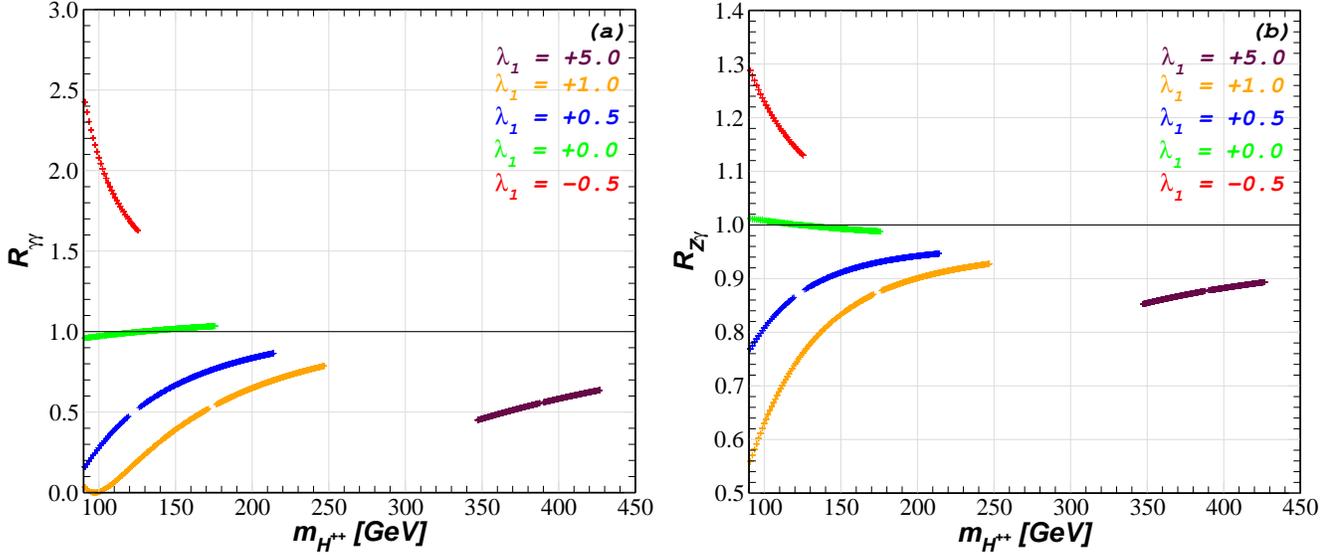

\hspace{-0.9cm}
\begin{tabular}{rr}
\hspace{-.8cm}\resizebox{86mm}{!}{\includegraphics{dgnr_Rgaga_mhpp.eps}}&
\hspace{.1cm}\resizebox{86mm}{!}{\includegraphics{dgnr_RgaZ_mhpp.eps}}
\end{tabular}
\vspace{0.1cm}
\caption{The $R_{\gamma\gamma}$ (left) and $R_{Z\gamma}$ (right) ratios as a function of $m_{H^{\pm\pm}}$ for various values of $\lambda_1$ with $v_t = 1$~GeV and $\le \lambda_2 \le 5$, $\lambda_3 = 2 \lambda_2$ and $-12 \le \lambda_4 \le 2$.}
\label{fig:Rxx_mhpp}
\end{figure}

\begin{figure}[ht]
\centering
\resizebox{86mm}{!}{\includegraphics{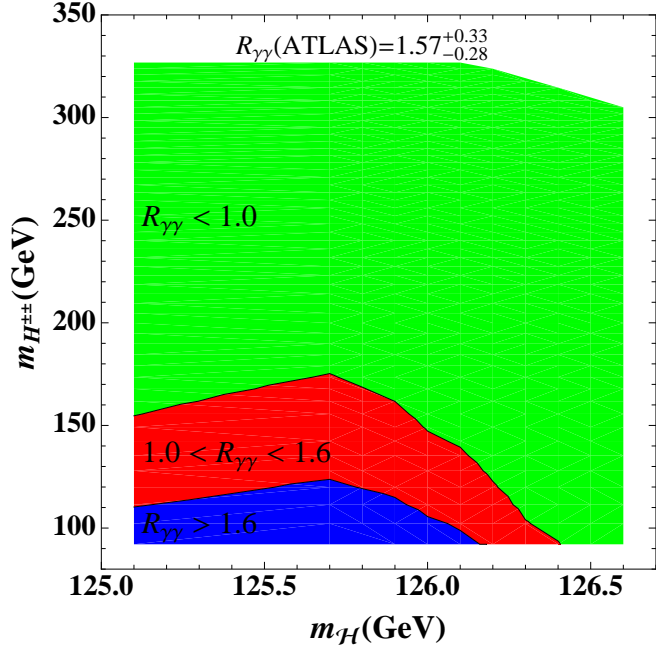}}
\caption{Scatter plots showing the $R_{\gamma\gamma}$ (left) and $R_{Z\gamma}$ (right) ratios in the $[ m_H, m_{H^{\pm\pm}}]$ with  $-1 \le \lambda_1 \le 3$, $-5 \le \lambda_2 \le 5$, $\lambda_3 = 2 \lambda_2$ and $-3 \le \lambda_4 \le 1$.}
\label{fig:contour_mhp_mhpp}
\end{figure}

Fig.~\ref{fig:Rxx_mhpp} illustrates the same ratios as a function of the doubly charged Higgs mass $m_{H^{\pm\pm}}$ for different values of $\lambda_1$. We see again that positive deviations from SM happen when the doubly charged Higgs gets a mass smaller than $180$~GeV with $\lambda_1$ in the vicinity of $0$ or negative. In this case, deviations of more than $200\%$ for $R_{\gamma\gamma}$ and $60\%$ for $R_{Z\gamma}$ are possible. This feature is clearly indicated in the contour plots of $R_{\gamma\gamma}$ in the ($m_{H}$, $m_{H^{\pm\pm}}$) in Fig.~\ref{fig:contour_mhp_mhpp}. which shows that $2\sigma$ consistency with ATLAS observed excess in diphoton channel constrains $m_{H^{\pm\pm}}$ to be in the range $92 - 180$~GeV.\\
 
\begin{figure}[ht]
\hspace{-1.cm}
\begin{tabular}{rr}
\hspace{-.8cm}\resizebox{86mm}{!}{\includegraphics{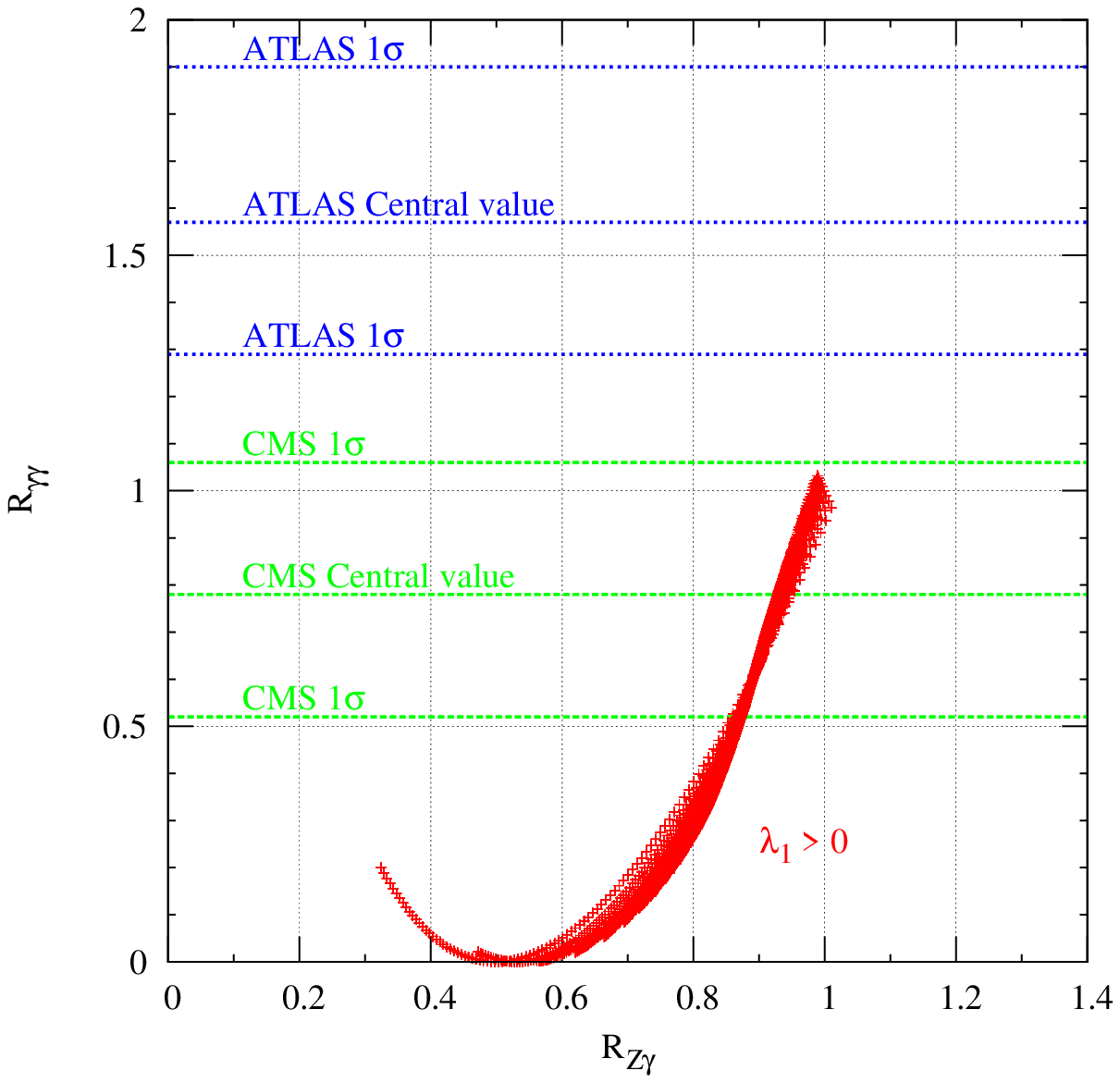}}&
\hspace{.1cm}\resizebox{86mm}{!}{\includegraphics{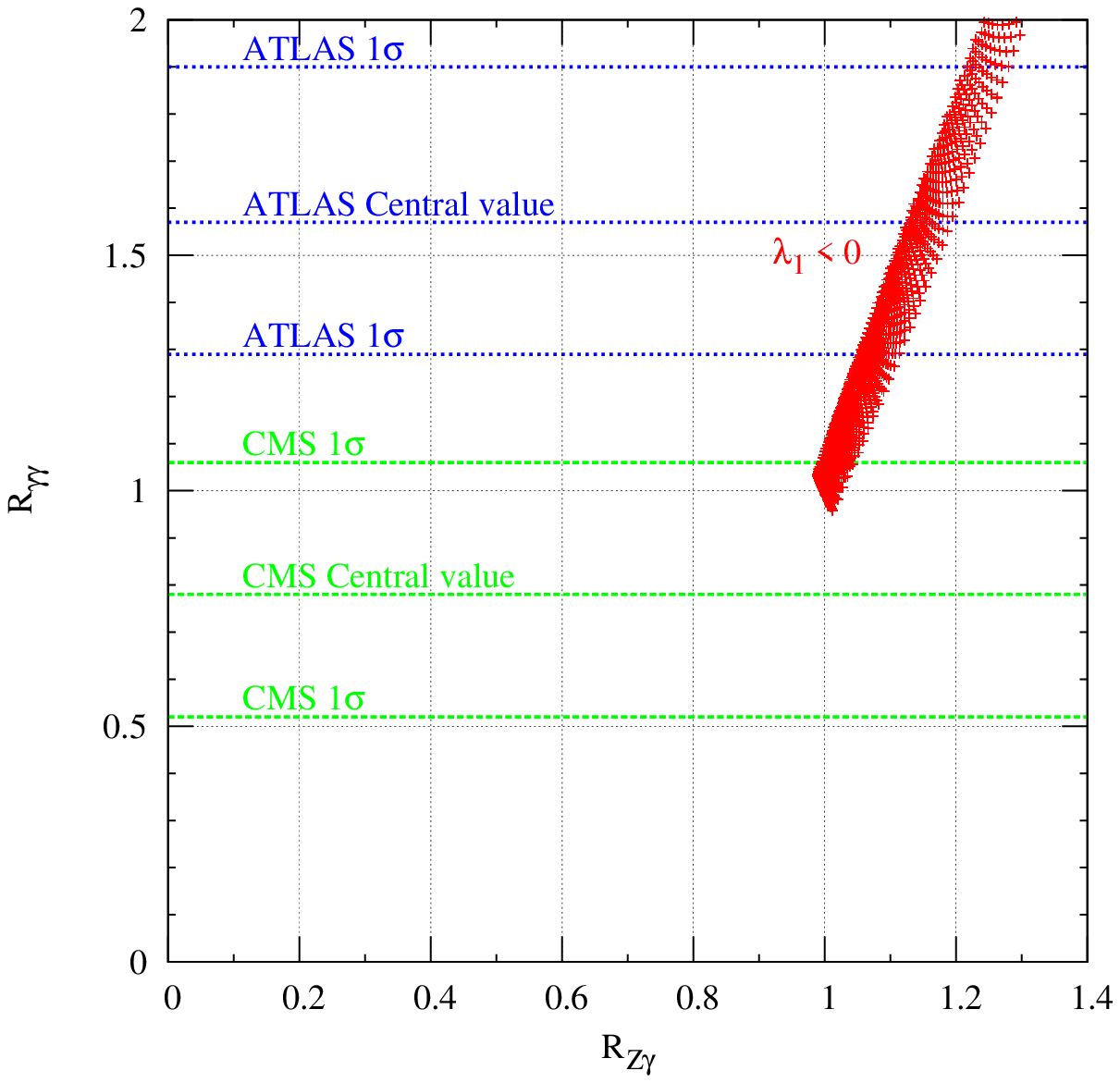}}\\
\end{tabular}
\caption{The $R_{\gamma\gamma}$ and $R_{Z\gamma}$ correlation with $\lambda = 0.53$, $-2 \le \lambda_1 \le 12$, $-5 \le \lambda_2 \le 5$, $\lambda_3 = 2 \lambda_2$ and $-12 \le \lambda_4 \le 2$.}
\label{fig:correlation}
\end{figure}

It is also interesting to comment on the correlation between the two channel $h\rightarrow\gamma\gamma$  and $h\rightarrow Z\gamma$. Fig.~\ref{fig:correlation} shows two main features: First, except a narrow region disfavoured by LHC measurements, these two decay modes are generally correlated. Second the correlation is very sensitive to the sign of the $\lambda_1$. \\

\begin{figure}[ht]
%\vspace{-2.8cm}
\hspace{-0.95cm}
\begin{tabular}{rr}
\hspace{-.8cm}\resizebox{86mm}{!}{\includegraphics{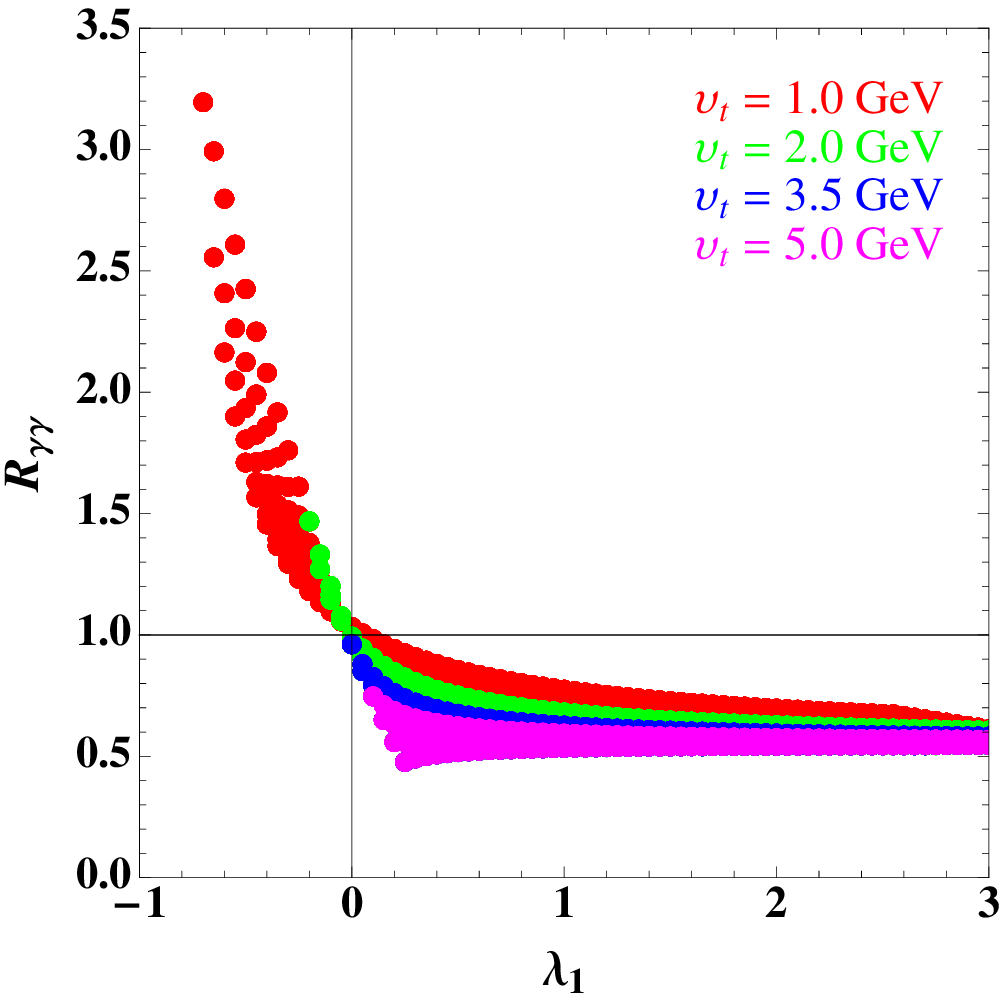}}&
\hspace{.1cm}\resizebox{86mm}{!}{\includegraphics{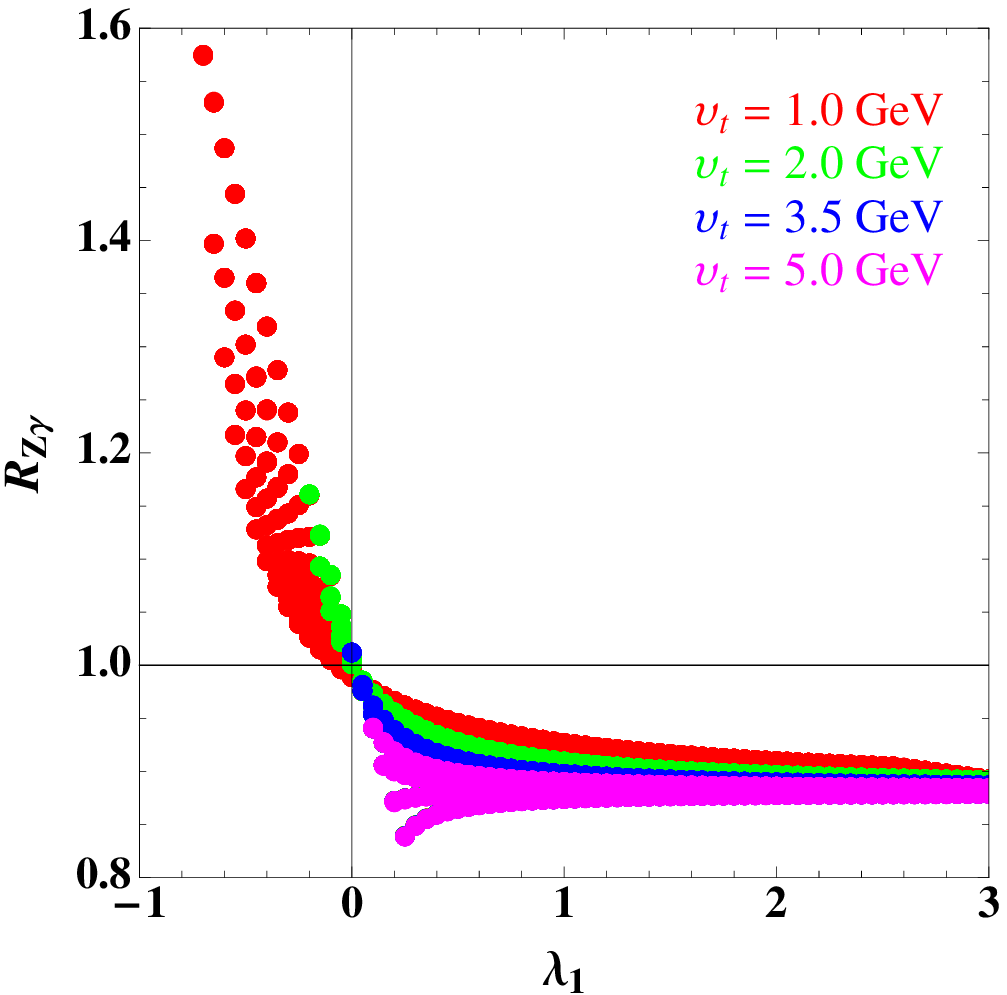}}
\end{tabular}
\caption{$R_{\gamma\gamma}$ (left) and $R_{Z\gamma}$ (right) ratios as a function of $\lambda_1$ for various values of $v_t$ with $\lambda = 0.530$, $-1 \le \lambda_1 \le 3$, $-5 \le \lambda_2 \le 5$, $\lambda_3 = 2 \lambda_2$ and $-3 \le \lambda_4 \le 1$.}
\label{fig:Rxx_ld1_diffvt}
\end{figure}

One last comment is in order. Let us see what happens when the triplet' VEV increases in this scenario. From Fig.~\ref{fig:Rxx_ld1_diffvt} we see that enhancements in both decay channels gradually reduce when $v_t$ becomes larger than $1$~GeV and disappear once $v_t$ exceeds $3.5$~GeV. 

%%%%%%%%%%%%%%%%%%%%%%%%%%%%%%%%%%%%%%%%%%%%%%%%%%%%%%%%%%%%%%%%%%%%%%%%%%%%%%%

\section{Conclusions}

\label{sec:conclusion}
The LHC has impressive sensitivity to Higgs bosons decays to two photons with a large branching
ratio. Observation of this photonic decay mode $(\mathcal{H}\rightarrow\gamma\gamma)$ with a rate
significantly above that expected for the Standard Model Higgs boson is a remarkable step toward
new physics. Also, the decay mode $\mathcal{H}\rightarrow Z\gamma$ is complementary to the diphoton channel and may provide additional valuable information on Higgs properties.   

We have calculated the branching ratio for these Higgs channels in the framework of the \textsc{htm} in the scenario where the two CP even Higgs are almost mass degenerate. Our analysis takes into account the full set of experimental and theoretical constraints including perturbative unitarity as well as vacuum stability constraints on the scalar potential parameters. .

In the Standard Model the decays $h\rightarrow\gamma\gamma$  and $h\rightarrow Z\gamma$ are
assumed to be mediated solely by loops involving $W^\pm$ and fermions.  In SM extension to HTM model, 
loop contributions are affected from charged scalar Higgses. Such contributions provide substantial enhancement 
for the $\gamma\gamma$ decay mode branching ratio, so $1< R_{\gamma\gamma}<3$, when $\lambda_1$ get rather small and negative values and $\lambda_4$ lies between $\frac{-1}{2}$ and $\frac{1}{4}$.  Also, we find that consistency with LHC signal strength favours a light doubly charged Higgs with a mass between $92$~GeV and $180$~GeV. As to $h\rightarrow Z \gamma$, we get a similar behaviour to the diphoton decay mode but with slight enhancement with ${\rm R}_{Z\gamma}$  not exceeding $1.6$.
Finally, we also study the correlation between the two decay modes in the mass degenerate scenario. The analysis shows two main features: 
First, these modes are strongly correlated in most of the allowed region and are quite sensitive to the sign of the $\lambda_1$ with $\lambda_1<0$
 favoured by LHC diphoton data.  

\section*{Acknowledgments}
M. C. would like to thank S. Randjbar-Daemi and F. Quevedo for the invitation and hospitality at the Abdus Salam ICTP Centre where part of this work has been done.  M. C. P. also thanks G. Moultaka (L2C/CNRS-Montpellier University) for  fruitful discussions. We are grateful to A. Arhrib for providing a code for Higgs decay into $Z \gamma$.  This work is supported in part by the Moroccan Ministry of Higher Education and Scientific Research: {\it projet des domaines prioritaires de la recherche scientifique et du developpement technologique.}

%--------------------------------------------
\newpage

\section*{Appendix: More on $\alpha$, BFB and unitarity constraints}
 The basic idea to know more about the mixing angle $\alpha$ is to use the BFB conditions.  Since $\lambda$ is already positive, we have to 
scrutinise the set of $6$ BFB constraints for the $4$ possible values of the mixing angle. First we note that some constraints 
depend on  $\sin^2{\alpha}$ and $\cos^2{\alpha}$ and are not sensitive to the sign of $\alpha$. This is true for the second and third constraints 
of Eq.~(\ref{eq:bound1}). It turns out that $\lambda_2 + \lambda_3$ is positive for $\alpha = \pm ( \pi/4 -2 v_t/v_d )$ in a large domain of $v_t$, and for 
the values of the other parameters as determined in subsection III-A. This contrasts with $\alpha = \pm ( \pi/4 +2 v_t/v_d )$ where the sign
 of $\lambda_2 + \lambda_3$ depends on the numerical value of $v_t$ in the range of interest. So to go further in an analytical way, we select 
$\alpha = \pm ( \pi/4 -2 v_t/v_d )$
as  good candidates, looking for a constraint able to disentangle these values. So we will turn on constraints depending on  $\lambda_1$ since this 
parameter depends on $\sin{2 \alpha}$. It appears that the first condition in Eq.~(\ref{eq:bound2}) is always positive if $\alpha = -\pi/4 +2 v_t/v_d $ and
if $ 2 m^2_{H^{\pm}} > m^2_{A_0}$, but, for $\alpha = \pi/4  - 2 v_t/v_d $, there are no simple conditions insuring the positivity. The last conditions 
of Eq.(\ref{eq:bound1}) is always true if $m_{H^0}^2 + m_{h^0}^2 + 2 m_{H^{\pm \pm}}^2 \geq 4 m_{H^{\pm}}^2$, for $\alpha = -\pi/4 +2 v_t/v_d $ 
(and $\alpha = \pi/4 -2 v_t/v_d $ as well). 
To sum up, all the BFB constraints are verified, in the degenerate scenario and with small $v_t/v_d$, if: 
                                            $$\alpha = - \pi/4 +2 v_t/v_d,$$ 
together with the mass relation:
\begin{equation} 
                             m_{H^0}^2 + m_{h^0}^2 + 2 m_{H^{\pm \pm}}^2 \geq 4 m_{H^{\pm}}^2 \geq  2 m^2_{A_0}
\label{eq:masses}
\end{equation}
We stress that they are not necessary conditions to ensure BFB conditions Eqs.~(\ref{eq:bound2}, \ref{eq:bound3}, \ref{eq:unit1}). The last inequality gives numerically
 $m_{H^{\pm}} > \frac{m_{A_0}}{\sqrt 2} \approx 88$~GeV.
 Given this mixing angle, we obviously see that $B$ is always positive, allowing to remove the absolute value in Eq.~(\ref{dgnr3}). Moreover, we also show that
$\nu= \displaystyle {\frac{m_{H^0}^2 - m_{h^0}^2}{2 v_t}} (v_d \cos 4 x + 4 v_t \sin 4 x )$ remains positive as long as $x =v_t/v_d \in
 [0,0.699]$, a large enough interval for our purpose. \\ 

Next, taking into account the previous conclusions, we focus on the unitarity constraints, Eqs.~(\ref{eq:unit1} - \ref{eq:unit10}), and discuss how they behave in nearly
 degenerate Higgs scenario. The unitarity requirement along with the set BFB conditions severely constrain the parameter space that one has to consider for phenomenological
 studies. Here it is noticeable that one gets nice simple formulas and Higgs mass bounds which may be very convenient to use. If we we assume that
 $v_d$,  $M_{ex}$ take their experimental values, roughly, $246$~GeV, $125$~GeV 
respectively and $\Delta M = 2$~GeV, we only have three unknown parameters to manage: $m_{H^{\pm}}$,  $m_{H^{\pm \pm}}$ and $v_t$, 
all GeV-dimensioned.\\

By keeping the leading terms (until linear in $v_t$), one can see that only $v_t$ is not fixed in Eqs.~$(\ref{eq:unit1}, \ref{eq:unit4}, \ref{eq:unit6}, \ref{eq:unit7}_{+})$. 
However, a careful study of those $4$ functions shows that the most efficient
 one to constrain $v_t$ comes from the unitarity bound equation Eq.~(\ref{eq:unit6}) where 
$v_t \geq \displaystyle {\frac{ \Delta M (2 + \sqrt{3+ \kappa \pi} ) }{2 (\kappa \pi -1)} }$~GeV. If  $\kappa = 8$, then $v_t \geq 0.30$~GeV.\\ 

The second step is to look at the formulas  Eqs.~(\ref{eq:unit2}, \ref{eq:unit3}, \ref{eq:unit9}) which depend both  on 
$v_t$ and $m_{H^{\pm}}$. These equations appear as sums of two single variable functions, one depends on $v_t$ while the other depends on $m_{H^{\pm}}$, 
a very useful property that allows to find upper bounds for  $m_{H^{\pm}}$. It turns out that we get the best one from Eq.~(\ref{eq:unit9}): 

 \begin{equation}
m_{H^{\pm}}^2 \leq \displaystyle{ \frac{\kappa \pi v_d^2 + 4 M_{ex}^2}{6} - \frac{\Delta M M_{ex} (v_d^2 - 6 v_t^2)}{ 6 v_d v_t}}
\label{eq:mhplowerbound}
\end{equation}
 For $\kappa=8$, the upper bound of $m_{H^{\pm}}$  found is $487$~GeV for $v_t = 0.30$~GeV, and increases to $511$~GeV for $v_t = 1$~GeV and to $519$~GeV for $v_t = 5$~GeV.\\

 Finally, the other five equations, namely Eqs.~$(\ref{eq:unit5}, \ref{eq:unit7}_{-}, \ref{eq:unit8}_{+}, \ref{eq:unit8}_{-}, \ref{eq:unit10})$\footnote{$\ref{eq:unit7}_{-}, \ref{eq:unit8}_{+}, \ref{eq:unit8}_{-}$ refer to Eq.~(\ref{eq:unit7}) and Eq.~(\ref{eq:unit8}) with ${-}$ or ${+}$ sign.}, contain $m_{H^{\pm \pm}}$ as well as $v_t$ and $m_{H^{\pm}}$. 
It appears that all these different equations give the same {\it numerical results in the allowed domain} of $v_t$ and $m_{H^{\pm}}$. 
Since these five equations yield equivalent results, we illustrate with Eq.~(\ref{eq:unit10}):
\begin{equation}
m_{H^{\pm}}^2 \geq \displaystyle{ \frac{{\Delta M}^2 + 12 M_{ex}^2}{24 } + \frac{ 2 \Delta M M_{ex} v_t}{3 v_d}}
\label{eq:mhpupperbound}
\end{equation}

\begin{equation} 
m_{H^{\pm \pm}}^2 \leq \displaystyle{ \frac{4 (M_{ex}^2 +\kappa \pi v_d^2) - {\Delta M}^2  }{12} - 
\frac{\Delta M M_{ex} (v_d^2 - 4 v_t^2)}{3 v_d v_t}}
\label{eq:mhpplowerbound}
\end{equation} 
\\ 
Numerically, we obtain $m_{H^{\pm}} \geq 88$~GeV as found before in Eq.~(\ref{eq:masses}),  which is compatible with the LEPII combined lower limit \cite{LEP} and agrees also with the bounds established by the LHC \cite{cms_charged, atlas_charged}. In the case of doubly charged Higgs, 
for $\kappa = 8$, we find $m_{H^{\pm \pm}} \leq 666$~GeV for $v_t = 0.30$~GeV, $m_{H^{\pm \pm}} \leq 701$~GeV for 
$v_t = 1$~GeV and $m_{H^{\pm \pm}} \leq 712$~GeV for $v_t = 5$~GeV. These mass predictions accommodate well the experimental 
upper limits $m_{H^{\pm \pm}}$ reported by the current search for multilepton final states performed by CMS  \cite{cms_dcharged} and ATLAS \cite{atlas_dcharged}.\\
    
Note that the redundancy in the last studied equations is a consequence that the number of combined equations involving together BFB 
and unitarity constraints is smaller than the sum of the numbers of equations for the two separate sets given by Eqs.~(\ref{eq:syscompact1}-\ref{eq:syscompact7}). 

.
%----------------------------

\newpage
%1
%----------------------------
%2
%3
%-------------------------------
%4
%----------------------------
%5
%----------------------------
%------------------------------------------------------------
%6
%---------------------------------------------------------
%----------------------------------------------------------------

\end{document}